\documentclass[journal, 10pt,twocolumn]{IEEEtran}
\usepackage{cite}
\usepackage{amsmath, amssymb, amsfonts, amsthm}
\usepackage{algorithm}
\usepackage[noend]{algorithmic}
\usepackage{graphicx}
\usepackage{textcomp}
\usepackage{xcolor}
\usepackage{color}
\usepackage{bm}
\usepackage{subfigure} 
\usepackage{booktabs}
\usepackage{autobreak}
\usepackage{multirow}
\usepackage{booktabs}
\usepackage{authblk}
\usepackage{mathrsfs}
\usepackage{array}
\usepackage{makecell} 
\usepackage{float}
\usepackage{soul}
\usepackage{stmaryrd}
\usepackage{dsfont}
\usepackage{bbm}
\usepackage[hidelinks]{hyperref}
\usepackage{cleveref}
\usepackage{pifont}
\usepackage{enumitem}
\usepackage{cancel}

\newtheorem{lemma}{Lemma}

\newtheorem{proposition}{Proposition}
\newtheorem{definition}{Definition}

\begin{document}
\title{On the Impact of Uncertainty and Calibration on Likelihood-Ratio Membership Inference Attacks}

\author{Meiyi Zhu, Caili Guo \IEEEmembership{Senior Member, IEEE}, Chunyan Feng \IEEEmembership{Senior Member, IEEE},\authorcr Osvaldo Simeone \IEEEmembership{Fellow, IEEE}

\vspace{-1cm}

\thanks{
The work of M. Zhu, C. Guo and C. Feng was supported by the National Natural Science Foundation of China (62371070), by the Fundamental Research Funds for the Central Universities (2021XD-A01-1), and by the Beijing Natural Science Foundation (L222043). The work of M. Zhu was also supported by the BUPT Excellent Ph.D. Students Foundation (CX2023150).
The work of O. Simeone was supported by the European Union's Horizon Europe project CENTRIC (101096379), by the Open Fellowships of the EPSRC (EP/W024101/1), and by the EPSRC project (EP/X011852/1).

Meiyi Zhu, Caili Guo and Chunyan Feng are with the Beijing Key Laboratory of Network System Architecture and Convergence, School of Information and Communication Engineering, Beijing University of Posts and Telecommunications, Beijing 100876, China (e-mail: lia@bupt.edu.cn; guocaili@bupt.edu.cn; cyfeng@bupt.edu.cn).

Osvaldo Simeone is with the King's Communications, Learning \& Information Processing (KCLIP) lab, Department of Engineering, King's College London, London WC2R 2LS, U.K. (e-mail: osvaldo.simeone@kcl.ac.uk).
}
}

\maketitle
\begin{abstract}
In a \textit{membership inference attack} (MIA), an attacker exploits the overconfidence exhibited by typical machine learning models to determine whether a specific data point was used to train a target model. In this paper, we analyze the performance of the {\textit{likelihood ratio attack} (LiRA)} within an information-theoretical framework that allows the investigation of the impact of the \textit{aleatoric uncertainty} in the true data generation process, of the \textit{epistemic uncertainty} caused by a limited training data set, and of the \textit{calibration level} of the target model. We compare three different settings, in which the attacker receives decreasingly informative feedback from the target model: \textit{confidence vector} (CV) disclosure, in which the output probability vector is released; \textit{true label confidence} (TLC) disclosure, in which only the probability assigned to the true label is made available by the model; and \textit{decision set} (DS) disclosure, in which an adaptive prediction set is produced as in conformal prediction. We derive bounds on the advantage of an MIA adversary with the aim of offering insights into the impact of uncertainty and calibration on the effectiveness of MIAs. Simulation results demonstrate that the derived analytical bounds predict well the effectiveness of MIAs.
\end{abstract}

\vspace{-4mm}

\vspace{-6mm}
\section{Introduction}
\subsection{Context and Motivation}
By attempting to infer whether individual data records were used to train a machine learning model, \textit{membership inference attacks} (MIAs) pose a significant privacy risk, particularly in sensitive domains such as healthcare or recommendation systems, revealing personal information without consent \cite{shokri2017membership, carlini2022membership}. Analyzing MIAs is crucial to understanding privacy threats to machine learning models, as well as to developing robust countermeasures \cite{yeom2018privacy, jayaraman2019evaluating,chen2023overconfidence}.

Machine learning models that exhibit \textit{overconfidence} in predictions are known to be more susceptible to MIAs \cite{shokri2017membership, carlini2022membership, chen2023overconfidence}. In fact, by assigning disproportionately high confidence scores to predictions on inputs present in the training set, overconfident models inadvertently leak information about their training data. This leakage enables attackers to infer the presence of specific data points in the training data set with greater accuracy, thereby compromising data privacy. Improving the \textit{calibration}, i.e., the uncertainty quantification capabilities, of predictive models is thus a common step to safeguard against MIAs \cite{chen2023overconfidence}.

However, the vulnerability to MIAs is not solely contingent upon model calibration. In fact, models with the same calibration level may exhibit significantly different susceptibilities to MIAs depending on the underlying predictive uncertainty level. Predictive uncertainty can be broadly classified into \textit{aleatoric uncertainty} and \textit{epistemic uncertainty} \cite{hullermeier2021aleatoric, guo2007sensitivity}. Aleatoric uncertainty reflects the inherent unpredictability of the data generation process; while epistemic uncertainty reflects limitations in the knowledge of the model caused by access to limited training data. As the predictive uncertainty increases, due to either aleatoric uncertainty or epistemic uncertainty, a model with a fixed calibration level would tend to produce more balanced confidence levels. The resulting uniformity in the model's behavior inside and outside the training set makes it more difficult to carry out a successful MIA \cite{jia2019memguard, nasr2018machine}.

Furthermore, an attacker without access to the target model weights is characterized by epistemic uncertainty, i.e., by uncertainty about the actual model parameters of the target model \cite{Liu2025efficient}. In fact, the variability in the model parameter space associated with an unknown training data set increases with smaller data set sizes, since larger data sets tend to all yield models that minimize the underlying population loss. 

Overall, exploring the impact and interplay of uncertainty and calibration is essential for a comprehensive understanding of MIAs, supporting the development of more effective privacy-preserving strategies.

\vspace{-3mm}
\subsection{State of the Art}
\textit{Confidence vector (CV) disclosure:} The key reference \cite{shokri2017membership} introduced MIAs by focusing on black-box scenarios in which the attacker can query the target model once for the given input of interest, obtaining the full \textit{confidence vector} (CV) produced at the output of the model. The authors turned the MIA into a classification problem, and trained a neural network to distinguish between the target model's responses to data points inside and outside the training set. To this end, the authors proposed to construct \textit{shadow models} that are designed to imitate the behavior of the target model for inputs inside and outside the training set. Using the shadow models, a neural network is trained to distinguish the two types of inputs. Follow-up work includes \cite{salem2018ml}, which relaxed the assumptions on the knowledge of data distribution and model architecture by leveraging an ensemble of basic classification networks. Attacks based on the evaluation of the entropy of the confidence probability vectors were studied in \cite{yeom2018privacy, song2021systematic}.

\textit{True label confidence (TLC) disclosure:} Based on the observation that overconfident models tend to generate higher confidence scores to the true label when queried on training data than on unseen test data \cite{chen2023overconfidence}, various studies explored settings in which the target model returns only the \textit{true label confidence} (TLC) level, or equivalently the corresponding prediction loss \cite{salem2018ml, yeom2018privacy, jayaraman2020revisiting, liu2022membership}. In such scenarios, the attacker sets a threshold on the target model's confidence assigned to the true label to identify the training data points \cite{yeom2018privacy, song2021systematic}. To optimize the threshold,  one can aim at maximizing the true positive rate (TPR) with a constraint on a maximum false positive rate (FPR) \cite{jayaraman2020revisiting,bertran2024scalable}.

Reference \cite{carlini2022membership} introduced a type of \textit{likelihood ratio attack} (LiRA), in which Gaussian distributions are fitted to the TLC generated by shadow models trained to imitate models trained with and without the candidate data point. LiRA is then conducted through a parametric likelihood ratio test. Reference \cite{ali2023membership} improved LiRA by adding adversarial noise to the input in order to amplify the output gap between the two classes of shadow models. Further enhancements include setting sample-specific thresholds \cite{ye2022enhanced} and the use of pairwise likelihood ratio tests calibrated through a Bayesian method \cite{zarifzadeh2023low}. We refer to this family of methods as LiRA-style attacks, which includes the original LiRA \cite{carlini2022membership} and its subsequent variants \cite{ali2023membership, ye2022enhanced, zarifzadeh2023low}, all of which follow the likelihood-ratio-based hypothesis testing paradigm using shadow models to estimate membership likelihoods.

\textit{Label-only and decision set (DS) disclosure:} As another disclosure model, other studies have developed \textit{label-only} attacks, for which the attacker has access solely to predicted labels. Reference \cite{yeom2018privacy} proposed to identify samples that are correctly classified as part of the training data sets \cite{chen2023overconfidence}. Further research exploited the fact that the predicted labels of overconfident models are less sensitive to perturbations on inputs derived from training data as compared to previously unseen data. By analyzing the variability of predicted labels in the presence of adversarial perturbations \cite{li2021membership} or data augmentations \cite{choquette2021label}, attackers can thus attempt to identify training data points.

In this paper, we generalize label-only disclosure to inference settings in which the target model generates a \textit{decision set} (DS), i.e., a subset of labels, as its prediction output. Decision sets are typically generated by including all the labels whose confidence is above a threshold. A DS provides information about the uncertainty of the prediction through its size, and it can be derived using formal methods such as conformal prediction \cite{vovk2005algorithmic, angelopoulos2021gentle, cohen2023calibrating}.

\textit{Theoretical analysis of membership inference attacks:}
Previous work has explored the impact of various factors on privacy risks via experiments \cite{shokri2017membership, carlini2022membership, chen2023overconfidence, truex2019demystifying, tonni2020data} or via scenario-specific theoretical analyses \cite{yeom2018privacy, jayaraman2019evaluating, jayaraman2020revisiting,del2023bounding, aubinais2023fundamental, kulynych2019disparate, long2018understanding}. While some studies have empirically identified factors such as data complexity, intra-class variance, number of classes, data entropy, model type, and model fairness as influential to MIA performance \cite{truex2019demystifying, tonni2020data}, these factors can ultimately be viewed as indirect manifestations of uncertainty and calibration. Given the multitude of such factors, our work aims to rigorously explore the fundamental impact of uncertainty and calibration on MIA performance through theoretical analysis.

Notably, a line of work is motivated by the opposite nature of the goal of MIAs -- revealing a training data point -- and of differential privacy (DP) mechanisms --  obscuring the presence or absence of a training data point \cite{dwork2014algorithmic}. Based on this observation, upper bounds on the attacker's advantage \cite{yeom2018privacy, jayaraman2019evaluating} and positive predictive value \cite{jayaraman2020revisiting} were studied for DP-protected models. These works focus on the theoretical relationship between the DP metrics and MIA success measures. Furthermore, reference \cite{del2023bounding} established universal bounds on the success rate of MIAs as a function of the target model's generalization gap, while assuming a Bayesian attacker. Results that target symmetric and redundancy-invariant algorithms were also discussed in \cite{aubinais2023fundamental}. Further theoretical analyses have explored the vulnerability to MIA across population subgroups \cite{kulynych2019disparate}, as well as the impact of generalization \cite{long2018understanding}.

Compared to these prior theoretical analyses, our work differs in four key aspects: 1) Most existing studies focus on TLC disclosure, typically in the form of losses \cite{yeom2018privacy, jayaraman2019evaluating, jayaraman2020revisiting, del2023bounding, kulynych2019disparate, long2018understanding}. In contrast, we systematically analyze three widely-used disclosure modes, i.e., CV, TLC, and DS, which have not been jointly studied in theoretical work. 2) Prior analyses adopt fundamentally different frameworks, such as DP-specific \cite{yeom2018privacy, jayaraman2019evaluating, jayaraman2020revisiting} and Bayesian \cite{del2023bounding} settings. Our framework is specifically designed for LiRA-style attacks \cite{carlini2022membership, ali2023membership, ye2022enhanced, zarifzadeh2023low}, a class of practical and efficient likelihood-ratio-based MIAs that lack theoretical attention. 3) While some previous works often aim to establish tight bounds \cite{yeom2018privacy, jayaraman2019evaluating, jayaraman2020revisiting}, our goal is to derive trend-aware bounds that reveal how key factors influence MIA performance, prioritizing interpretability over tightness. 4) Finally, prior studies focus on factors such as model architecture \cite{jayaraman2019evaluating}, data size \cite{aubinais2023fundamental} and training epochs \cite{long2018understanding}. We instead focus on deeper underlying factors, i.e., aleatoric uncertainty, epistemic uncertainty, and calibration error, which, as shown in Appendix \ref{apdx_other_factor}, explain and unify many lower-level effects observed in prior work \cite{shokri2017membership, salem2018ml, truex2019demystifying, kulynych2019disparate}.

\vspace{-0.3cm}
\subsection{Contributions and Organization}
\begin{figure}[t]
    \centering
    \setlength{\abovecaptionskip}{-2pt}
    {\includegraphics[width = 0.49\textwidth]{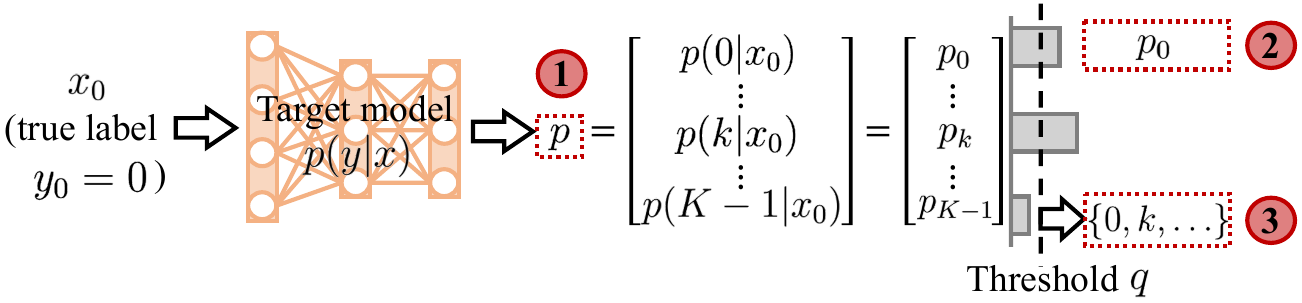}}
    \caption{Membership inference attack (MIA): An attacker queries an unknown target $K$-class classification model with some input $x_0$ whose true label, known to the attacker, is $y_0=0$ in order to determine whether data point $(x_0,y_0)$ was used in the training of the model. The attacker receives from the model one of the following: \textcircled{\scriptsize{1}} the \textit{confidence vector} (CV) $p$; \textcircled{\scriptsize{2}} the \textit{true label confidence} (TLC) $p_0 = p(0|x_0)$; or \textcircled{\scriptsize{3}} the \textit{decision set} (DS) $O(p)$ consisting of all labels $k$ with confidence level $p(k|x_0)$ no smaller than a threshold $q$.}\label{diff_observation}
    \vspace{-5mm}
\end{figure}

In this paper, we introduce a unified theoretical framework to analyze the performance of MIA under CV, TLC, and DS disclosure modes, assuming that the attacker follows the membership inference game studied in \cite{yeom2018privacy, jayaraman2020revisiting, carlini2022membership, ye2022enhanced, zarifzadeh2023low} (see Definition \ref{def_MIG} in Sec. \ref{testing}). Rather than proposing a new attack, our objective is to provide a theoretical analysis of existing LiRA-style attacks \cite{carlini2022membership, ali2023membership, ye2022enhanced, zarifzadeh2023low}, focusing on the impact and interplay of calibration and predictive uncertainty. The main contributions of this paper are summarized as follows.
\begin{itemize}
  \item As illustrated in Fig. \ref{diff_observation}, treating MIA design as a hypothesis-testing problem, we introduce a general information-theoretical framework to analyze the attacker's advantage for CV, TLC, and DS disclosure settings. Specifically, as seen in Fig. \ref{LiRA_Gauss}, our methodology analyzes LiRA-style attacks \cite{carlini2022membership, ali2023membership, ye2022enhanced, zarifzadeh2023low} for a wide range of disclosure settings, beyond the case of TLC disclosure considered in prior work. LiRA-style attacks use shadow models to imitate the behavior of the target model for inputs inside and outside the training set.
  \item The proposed framework models the confidence probability vectors produced by the shadow models using Dirichlet distributions with the aim of accounting for both aleatoric uncertainty -- the inherent randomness in the data-generation process -- and epistemic uncertainty -- quantifying the variability of the shadow models as a function of the unknown training set. Under this working assumption, information-theoretic upper bounds on the attacker's advantage under CV, TLC, and DS observations are derived that enable a theoretical investigation of the effectiveness of MIA  as a function of the model's calibration, as well as of the aleatoric and epistemic uncertainties.
  \item Simulation results validate the analysis, providing insights into the influence of calibration and predictive uncertainty on the attacker's advantage under CV, TLC, and DS disclosure settings.
\end{itemize}

The remainder of this paper is organized as follows. In Sec. \ref{problem_definition}, we outline the problem, review the LiRA framework, and introduce relevant performance metrics. Sec. \ref{preli_definitions} reviews a performance bound for MIA, and defines aleatoric uncertainty and calibration. In Sec. \ref{analysis}, we analyze the performance of MIA given CV, TLC, and DS observations. Sec. \ref{experiments} presents simulation-based results to evaluate the performance of MIA, validating our theoretical findings. Sec. \ref{LiRA_conclusion} concludes the paper.

\vspace{-3mm}
\section{Problem Definition} \label{problem_definition}

\subsection{Setting}\label{testing}
We focus on a $K$-class classification problem in which the target model belongs to a parameterized model class
\begin{align}\label{eq_mod_cla}
    \mathcal{F} = \left\{p(y|x,\theta):\theta\in\mathbb{R}^d\right\},
\end{align}
consisting of conditional probability distributions over the label variable $y\in\left\{0,1,\ldots,K-1\right\}$ given input $x$ that is parameterized by a vector $\theta$. Let $\mathcal{T}(\cdot)$ denote a \textit{training algorithm} that may incorporate any randomness arising from stochastic steps such as mini-batch selection, and $p^*(x,y)$ be the \textit{ground-truth data distribution}. As in prior art \cite{yeom2018privacy, jayaraman2020revisiting, carlini2022membership, ye2022enhanced, zarifzadeh2023low}, we define the \textit{membership inference game} as follows.

\begin{definition}[\textbf{Membership Inference Game}]\label{def_MIG}
    A membership inference game consists of the following steps:
    \begin{itemize}
        \item[{\textnormal{1)}}] A training data set $\mathcal{D}^{\mathrm{tr}}$ is generated by sampling data points in an independent and identically distributed (i.i.d.) way from the ground-truth distribution $p^*(x,y)$. This training data set is used to train the target model $p(y|x,\theta_{\mathcal{D}^{\mathrm{tr}}})$, with trained parameters obtained from the training algorithm as $ \theta_{\mathcal{D}^\mathrm{tr}}= \mathcal{T}(\mathcal{D}^{\mathrm{tr}})$.
        \item[{\textnormal{2)}}] A bit $b\in\left\{0,1\right\}$ is drawn uniformly at random. If $b=1$, a data point $(x_0,y_0)$ is selected from the training set $\mathcal{D}^{\mathrm{tr}}$, while, if $b=0$, a new data point $(x_0,y_0)\sim p^*(x,y)$ is generated from the ground-truth distribution in a manner independent of the training set.
        \item[{\textnormal{3)}}]Given the pair $(x_0,y_0)$, the attacker estimates the binary variable $b$ by leveraging its knowledge of the ground-truth distribution $p^*(x,y)$, of the model class $\mathcal{F}$, and of the training algorithm $\mathcal{T}(\cdot)$.
    \end{itemize}
\end{definition}
This game makes the worst-case assumption that the adversary knows the data generation distribution $p^*(x,y)$, so that it can generate data from it. In practice, this situation is approximately valid as many attacks require access to samples from the ground-truth distribution \cite{shokri2017membership, yeom2018privacy, jayaraman2020revisiting, salem2018ml, carlini2022membership, ye2022enhanced, zarifzadeh2023low, truex2019demystifying}. From these samples, the data generation distribution can be estimated using techniques such as model-based synthesis and statistics-based synthesis \cite{shokri2017membership}. The game also posits that the attacker can use the same model architecture $\mathcal{F}$ and training algorithm $\mathcal{T}(\cdot)$ as the target model. This may be the case if the adversary can leverage the same machine learning-as-a-service provider that builds the target model \cite{shokri2017membership, salem2018ml}, or if it performs model extraction to approximate the target model's behavior \cite{wang2018stealing}. Importantly, the attacker does not have access to the target model's training data set $\mathcal{D}^{\mathrm{tr}}$ or to its trained parameter vector $\theta_{\mathcal{D}^{\mathrm{tr}}}$.

\vspace{-0.3cm}
\subsection{Likelihood Ratio Attack}
In this paper we study a \textit{membership inference attack} (MIA) that follows the assumptions of the membership inference game in Definition \ref{def_MIG}. The attacker observes the output of the target model for the given input $x_0$. The information released by the target model is a function of the confidence probability vector $p(y|x_0,\theta_{\mathcal{D}^{\mathrm{tr}}})$ output by the model in response to a query with input $x_0$. We write the \textit{confidence probability vector} as
\begin{equation}\label{entire_v}
    p=
    \begin{bmatrix}
        p_0 \\
        \vdots \\
        p_{K-1}
    \end{bmatrix}
    =
    \begin{bmatrix}
        p(y=0|x_0) \\
        \vdots \\
        p(y=K-1|x_0)
    \end{bmatrix}
\end{equation}
with $p_k\in[0,1]$ and $\sum_{k=0}^{K-1} p_k=1$.
Without loss of generality, we reorder the entries so that the first element, $p_0$, corresponds to the probability assigned by the model to the true label $y_0$.

As illustrated in Fig. \ref{diff_observation}, we consider the following three different query outputs: \normalsize{\textcircled{\footnotesize{1}}} \textit{confidence vector} (CV) \cite{shokri2017membership}, which produces the entire vector \eqref{entire_v} as its output, i.e.,
\begin{equation}\label{obs_p}
    O(p)=p;
\end{equation}
\normalsize{\textcircled{\footnotesize{2}}} \textit{true label confidence} (TLC) \cite{carlini2022membership}, which only returns the confidence of the true label, i.e.,
\begin{equation}\label{obs_p0}
    O(p)=p_0;
\end{equation}
\normalsize{\textcircled{\footnotesize{3}}} \textit{decision set} (DS) \cite{angelopoulos2021gentle}, which generates a set of labels whose corresponding confidence levels are no smaller than a given threshold $q$.
\begin{equation}\label{obs_q_p}
    O(p)=\mathds{1}(p\geq q),
\end{equation} where $\mathds{1}(\cdot)$ is the indicator function and the threshold $q$ is known to the attacker.

A CV observation \eqref{obs_p} provides the attacker with the most information about the model, and it can thus be considered as the worst-case setting for the model vis-\`a-vis the attacker \cite{shokri2017membership}. The model may, however, only release the confidence level for the chosen label $y_0=0$, as studied in \cite{carlini2022membership}.
The resulting TLC output \eqref{obs_p0} generally results in less effective attacks as compared to CV observations. Finally, a DS output \eqref{obs_q_p} is common for models designed for reliable decision-making using tools such as \textit{conformal prediction} \cite{vovk2005algorithmic}. An attack based on \eqref{obs_q_p} may be considered as a form of \textit{label-only} attack \cite{choquette2021label}, which has not been explicitly studied in the literature.

Based on the output $O(p)$ of the model \eqref{obs_p}, \eqref{obs_p0}, or \eqref{obs_q_p}, in LiRA, the attacker aims at distinguishing two hypotheses\begin{subequations}\label{test}
    \begin{align}
        &\mathcal{H}^{\mathrm{out}}: (x_0,y_0)\notin\mathcal{D}^{\mathrm{tr}},\\
        &\mathcal{H}^{\mathrm{in}}: (x_0,y_0)\in\mathcal{D}^{\mathrm{tr}}.
    \end{align}
\end{subequations}Accordingly, the null hypothesis $\mathcal{H}^{\mathrm{out}}$ posits that the target model $p(y|x,\theta_{\mathcal{D}^{\mathrm{tr}}})$ is trained on a data set $\mathcal{D}^{\mathrm{tr}}$ that does not include the target example $(x_0,y_0)$, while the alternative hypothesis $\mathcal{H}^{\mathrm{in}}$ asserts the opposite.

Based on the observation $O=O(p)$, the attacker computes a test function $T(O)$, and then it decides for either hypothesis according to the rule\begin{subequations}\label{LiRA_hypothesis}
    \begin{align}
        \mathrm{if}&~T(O)=0,~\mathrm{choose}~\mathcal{H}^{\mathrm{out}},\\
        \mathrm{else~if} &~ T(O)=1, ~\mathrm{choose} ~\mathcal{H}^{\mathrm{in}}.
    \end{align}
\end{subequations}

\begin{figure}[t]
    \centering
    \setlength{\abovecaptionskip}{-2pt}
    {
	\includegraphics[width = 0.33\textwidth]{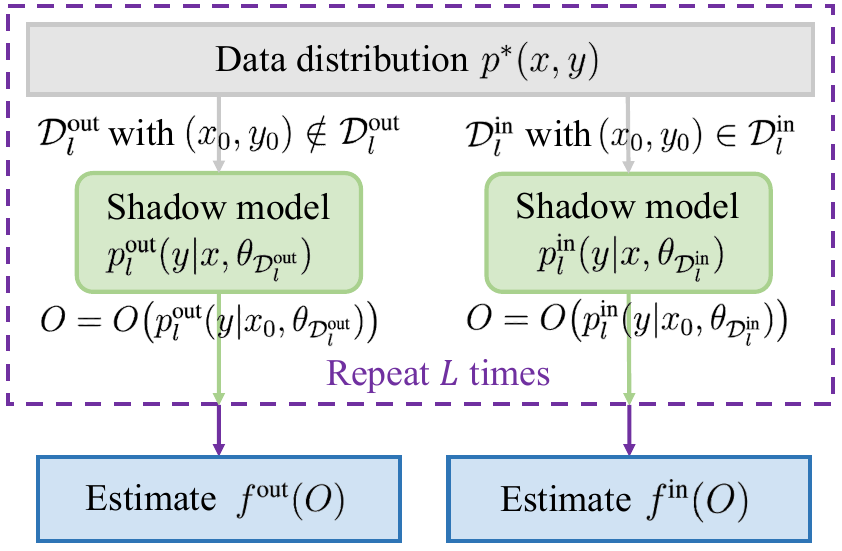}
	}
    \caption{Illustration of the estimation process of the distribution $f^{\mathrm{out}}(O)$ of the observation $O$ under the assumption $\mathcal{H}^{\mathrm{out}}$ that the training data does not include $(x_0,y_0)$ and of the distribution $f^{\mathrm{in}}(O)$ under the assumption $\mathcal{H}^{\mathrm{in}}$ that the target model is trained using the pair $(x_0,y_0)$.} \label{LiRA_Gauss}
    \vspace{-5mm}
\end{figure}

In order to construct the test variable $T(O)$, in this paper, we focus on the likelihood ratio attack (LiRA) \cite{carlini2022membership}, which follows the membership inference game in Definition \ref{def_MIG}. We refer to Sec. \ref{LiRA_conclusion} for some discussions on LiRA-style attacks with less powerful adversaries \cite{ali2023membership, ye2022enhanced, zarifzadeh2023low}. As illustrated in Fig. \ref{LiRA_Gauss}, LiRA estimates the distributions $f^{\mathrm{out}}(O)$ and $f^{\mathrm{in}}(O)$ of the observation $O$ during a preliminary phase. It is emphasized that the original LiRA focused solely on TLC outputs \eqref{obs_p0}, and assumed Gaussian distributions $f^{\mathrm{out}}(O)$ and $f^{\mathrm{in}}(O)$ to obtain a practical algorithm. In this paper, we study a more general framework, allowing for any of the outputs $O$ in \eqref{obs_p}--\eqref{obs_q_p}, with the aim of deriving analytical insights.

Using the available information about the distribution $p^*(x,y)$, training algorithm $\mathcal{T}(\cdot)$, and model architecture $p(y|x,\theta)$, in LiRA, the attacker constructs $L$ independent \textit{pairs of shadow models}, with each shadow model sharing the same architecture and training algorithm as the target model. To train each model pair $l=1, 2,\ldots,L$, the attacker generates two data sets. The first data set, $\mathcal{D}_l^{\mathrm{out}}$, contains $N^{\mathrm{tr}}$ examples sampled i.i.d. from the data distribution $p^*(x,y)$; while the second data set, $\mathcal{D}_l^{\mathrm{in}}$, contains $N^{\mathrm{tr}}-1$ examples sampled i.i.d. from $p^*(x,y)$ as well as the target input-output pair $(x_0,y_0)$:
\begin{subequations}
    \begin{align}
        &\mathcal{D}^{\mathrm{out}}_l=\left\{(x_n,y_n)\right\}_{n=1}^{N^{\mathrm{tr}}}~\mathrm{with}~(x_n,y_n)\sim p^*(x,y)\label{D_out}\\
        &\mathcal{D}^{\mathrm{in}}_l=\left\{(x_0,y_0)\right\}\cup\left\{(x_n,y_n)\right\}_{n=2}^{N^{\mathrm{tr}}}~\mathrm{with}~(x_n,y_n)\sim p^*(x,y)\label{D_in}
    \end{align}
\end{subequations}
In practice, the sets \eqref{D_out}--\eqref{D_in} can be obtained using data related to the training data set.

As illustrated in Fig. \ref{LiRA_Gauss}, using the data sets $\mathcal{D}_l^{\mathrm{out}}$ and $\mathcal{D}_l^{\mathrm{in}}$ and the known training algorithm $\mathcal{T}(\cdot)$, the attacker trains two models as 
\begin{align}
    \theta_{\mathcal{D}_l^{\mathrm{out}}}=\mathcal{T}(\mathcal{D}_l^{\mathrm{out}})~\mathrm{and}~\theta_{\mathcal{D}_l^{\mathrm{in}}}=\mathcal{T}(\mathcal{D}_l^{\mathrm{in}})
\end{align}
for all $l=1,2,\ldots,L$. Using the $L$ models $\{p(y | x, \theta_{\mathcal{D}_l^{\mathrm{out}}})\}_{l = 1}^L$, the attacker estimates the distribution $f^{\mathrm{out}}(O)$ of the outputs $O$ in \eqref{obs_p}--\eqref{obs_q_p} under the hypothesis $\mathcal{H}^{\mathrm{out}}$; while distribution $f^{\mathrm{in}}(O)$ of the outputs $O$ under the hypothesis $\mathcal{H}^{\mathrm{in}}$ is similarly estimated from the $L$ models $\{p(y|x, \theta_{\mathcal{D}_l^{\mathrm{in}}})\}_{l=1}^L$.

Distributions $f^{\mathrm{out}}(O)$ and $f^{\mathrm{in}}(O)$ represent the probability density functions of the observations $O$ under the hypotheses $\mathcal{H}^{\mathrm{out}}$ and $\mathcal{H}^{\mathrm{in}}$, respectively. Accordingly, the randomness captured by the distributions $f^{\mathrm{out}}(O)$ and $f^{\mathrm{in}}(O)$ reflects the inherent uncertainty of the attacker about the training data set $\mathcal{D}^{\mathrm{tr}}$ used to optimize the target model using algorithm $\mathcal{T}(\cdot)$. For the analysis in what follows, we assume $L$ to be large enough that distributions $f^{\mathrm{out}}(O)$ and $f^{\mathrm{in}}(O)$ are correctly estimated by the attacker. This assumption represents a meaningful worst-case condition for the target model, since $L$ is only limited by the computational complexity of the attacker.

Once the distributions $f^{\mathrm{out}}(O)$ and $f^{\mathrm{in}}(O)$ are estimated for the given pair $(x_0,y_0)$ of interest, LiRA applies a likelihood ratio test to the output $O$ produced by the target model in response to input $x_0$. Accordingly, the decision variable used in the decision \eqref{LiRA_hypothesis} is given by
\begin{equation}\label{LiRA_test}
    T(O)=\mathds{1}\Big(\log\Big(\frac{f^{\mathrm{out}}(O)}{f^{\mathrm{in}}(O)}\Big)< \tau\Big)
\end{equation}
for some threshold $\tau\geq 0$.

\vspace{-0.3cm}
\subsection{Performance Metrics}
The effectiveness of the test \eqref{LiRA_test}, and hence of the LiRA, is gauged by the \textit{type-\uppercase\expandafter{\romannumeral1} error} $1-\alpha$ and the \textit{type-\uppercase\expandafter{\romannumeral2} error} $\beta$. The parameter $\alpha$ represents the \textit{true negative rate} (TNR), i.e., the probability of correctly reporting that sample $(x_0,y_0)$ was not used to train the target model. The type-\uppercase\expandafter{\romannumeral2} error $\beta$ is, conversely, the probability of the attacker failing to detect that sample $(x_0,y_0)$ was not used for training, which is also referred to as the \textit{false negative rate} (FNR).

Mathematically, the TNR $\alpha$ and the FNR $\beta$ are defined as\begin{subequations}\label{def_alpha_beta}
    \begin{align}
    \alpha&=\mathbb{E}_{f^{\mathrm{out}}}\left[T(O)=0\right],\\
        \mathrm{and}~\beta&=\mathbb{E}_{f^{\mathrm{in}}}\left[T(O)=0\right],
    \end{align}
\end{subequations}
where $\mathbb{E}_{f^{\mathrm{out}}}[\cdot]$ and $\mathbb{E}_{f^{\mathrm{in}}}[\cdot]$ represent the expectations over the distributions $f^{\mathrm{out}}(O)$ and $f^{\mathrm{in}}(O)$, respectively, and the test variable $T(O)$ is defined in \eqref{LiRA_test}. Note that a larger TNR $\alpha$ and a smaller FNR $\beta$ indicate a more successful MIA.

\begin{figure}[t]
    \centering
    \setlength{\abovecaptionskip}{-2pt}
    {
	\includegraphics[width = 0.22\textwidth]{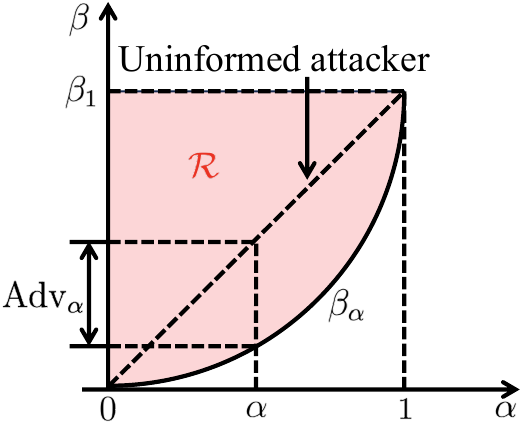}
	}
    \caption{Illustration of the NP region $\mathcal{R}$ (shaded region), of the trade-off function $\beta_{\alpha}$, and of the MIA advantage with respect to an uninformed attacker.}\label{Fig_NP_region}
    \vspace{-5mm}
\end{figure}

As illustrated in Fig. \ref{Fig_NP_region}, the \textit{Neyman-Pearson (NP) region} $\mathcal{R}$ consists of all the achievable pairs $(\alpha,\beta)$ of TNR and FNR values that can be attained by test \eqref{LiRA_test} when varying the threshold $\tau$. More precisely, including in region $\mathcal{R}$ all pairs $(\alpha',\beta')$ that are worse than a pair $(\alpha,\beta)$ obtained as per \eqref{def_alpha_beta} for some threshold $\tau$, we have
\begin{align}\label{NP_region}
    \mathcal{R}=&\big\{(\alpha', \beta')\in[0,1]^2: \exists\tau\geq 0\nonumber\\
    &\mathrm{s.t.}~\alpha'\leq \alpha~\mathrm{and}~\beta'\geq\beta~\mathrm{for}~(\alpha,\beta)~\mathrm{in}~\eqref{def_alpha_beta} \big\}.
\end{align}
Note, in fact, that any pair $(\alpha',\beta')$ with TNR $\alpha'\leq\alpha$ and FNR $\beta'\geq\beta$ is less desirable to the attacker than the pair $(\alpha,\beta)$.

The minimum FNR $\beta$ given a target TNR $\alpha$ is given by the \textit{trade-off function}
\begin{align}\label{tradeoff_function}
    \beta_{\alpha} = \inf_{(\alpha,\beta)\in\mathcal{R}}\beta.
\end{align}
For any fixed target TNR $\alpha$, a smaller trade-off function $\beta_{\alpha}$ indicates a more effective MIA. An \textit{uninformed attacker} that chooses hypothesis $\mathcal{H}^{\mathrm{out}}$ with probability $\alpha$, irrespective of the output $O$, achieves the trivial trade-off $\beta_{\alpha} = \alpha$. Consequently, an effective attack must satisfy the inequality $\beta_{\alpha} < \alpha$ for at least some values of the TNR $\alpha$.

Accordingly, to characterize how well an adversary can distinguish between the two hypotheses $\mathcal{H}^{\mathrm{out}}$ and $\mathcal{H}^{\mathrm{in}}$ in \eqref{test}, we introduce the MIA \textit{advantage} as the difference
\begin{align}\label{eq_Adv}
    \mathrm{Adv}_{\alpha} = \alpha - \beta_{\alpha}.
\end{align}
A larger advantage $\mathrm{Adv}_{\alpha}$ indicates a more successful MIA at TNR level $\alpha$, with $\mathrm{Adv}_{\alpha}=0$ corresponding to the performance of an uninformed attacker, which produces random guesses.

\vspace{-0.3cm}
\section{Preliminaries and Definitions}\label{preli_definitions}
In this section, we introduce relevant background information, as well as key definitions for the analysis in the next section.

\vspace{-0.3cm}
\subsection{Preliminaries}
By \eqref{LiRA_test} and \eqref{def_alpha_beta}, for a fixed TNR $\alpha$, the threshold $\tau_{\alpha}\in\mathbb{R}$ is uniquely determined by the equation \cite{murphy2012}
\begin{align}\label{alpha_LRT}
    \alpha=\mathbb{E}_{f^{\mathrm{out}}} \Big[\mathds{1}\Big(\log \Big(\frac{f^{\mathrm{out}}\left(O\right)}{f^{\mathrm{in}}\left(O\right)}\Big) \geq \tau_{\alpha}\Big)\Big],
\end{align}
and the trade-off function $\beta_{\alpha}$ can be accordingly calculated as
\begin{align}\label{beta_LRT}
    \beta_{\alpha}=\mathbb{E}_{f^{\mathrm{in}}} \Big[\mathds{1}\Big(\log\Big( \frac{f^{\mathrm{out}}\left(O\right)}{f^{\mathrm{in}}\left(O\right)}\Big) \geq \tau_{\alpha}\Big)\Big].
\end{align}
The characterization \eqref{alpha_LRT}--\eqref{beta_LRT} of the trade-off function $\beta_{\alpha}$ generally requires the numerical evaluation of the threshold $\tau_{\alpha}$, making it difficult to obtain analytical insights. A more convenient partial characterization is provided by the following information-theoretic outer bound.
\begin{lemma}[\textbf{Theorem 14.7 \cite{murphy2012}}]\label{lemma_outer_bound}
    Each pair $(\alpha,\beta)\in\mathcal{R}$ in the NP region \eqref{NP_region} satisfies the inequalities
    \begin{subequations}\label{outer_bound}
        \begin{align}
            &d(\alpha\|\beta)\leq D\left(f^{\mathrm{\rm{out}}}\|f^{\mathrm{\rm{in}}}\right),\\
            \mathrm{and}~&d(\beta\|\alpha)\leq D\left(f^{\mathrm{\rm{in}}}\|f^{\mathrm{\rm{out}}}\right),
        \end{align}
    \end{subequations}
    where
    \begin{align}
        D(p\|q)=\mathbb{E}_{p(x)}\Big[\log\frac{p(x)}{q(x)}\Big]
    \end{align}    
    is the Kullback-Leibler (KL) divergence between distributions $p(x)$ and $q(x)$, and
    \begin{align}
        d(a\|b)=a\log\Big(\frac{a}{b}\Big)+(1-a)\log\Big(\frac{1-a}{1-b}\Big)
    \end{align}
    is the binary KL divergence.
\end{lemma}

Lemma \ref{lemma_outer_bound} establishes a necessary condition for identifying the NP region in Fig. \ref{Fig_NP_region}. Consequently, the set of all $(\alpha,\beta)$ pairs satisfying the inequalities \eqref{outer_bound} encompasses the NP region. Thus, the function
\begin{equation}\label{lower_bound1}
    \beta_{\alpha}^{\mathrm{lb}}=\inf_{(\alpha,\beta):~ \eqref{outer_bound}~\mathrm{holds}}\beta
\end{equation}
serves as a lower bound on the optimal trade-off function \eqref{tradeoff_function}, i.e., we have the inequality
\begin{equation}\label{lower_bound}
    \beta_{\alpha}\geq\beta_{\alpha}^{\mathrm{lb}}
\end{equation}
for all $\alpha\in[0,1]$.

The following result follows from Pinsker's inequality \cite{jobic2023federated}, which yields the inequalities $d(\alpha\|\beta)\geq (\alpha-\beta)^2/2$ and $d(\beta\|\alpha)\geq (\beta-\alpha)^2/2$.
\begin{lemma}\label{lemma_ub}
    The advantage \eqref{eq_Adv} satisfies the upper bound
    \begin{align}\label{lower_bound2}
        \mathrm{Adv}_{\alpha}\leq \alpha - \beta_{\alpha}^{\mathrm{lb}}\leq \sqrt{D\left(f^{\mathrm{\rm{out}}} \| f^{\mathrm{\rm{in}}}\right) + D\left(f^{\mathrm{\rm{in}}} \| f^{\mathrm{\rm{out}}}\right)},
    \end{align}
    where $\beta_{\alpha}^{\mathrm{\rm{lb}}}$ is defined in \eqref{lower_bound1}.
\end{lemma}

\vspace{-0.3cm}
\subsection{Aleatoric Uncertainty}\label{CE_AU}
Let us write as $p^*_0=p^*(y=0|x_0)$ the \textit{ground-truth} probability of the true label $y_0$. Accordingly, the complementary probability
\begin{equation}\label{AU}
    \epsilon_a = 1-p^*_0\in\Big[0,1-\frac{1}{K}\Big]
\end{equation}
can be considered to be a measure of the irreducible uncertainty about the true label $y_0$ for the given input $x_0$. This is also known as \textit{aleatoric uncertainty}, or \textit{data uncertainty} \cite{hullermeier2021aleatoric}. The aleatoric uncertainty $\epsilon_a$ in \eqref{AU} is minimal, i.e., $\epsilon_a=0$, when the true label $y_0$ is certain, i.e., when we have $p_0^*=1$; and it is maximal, i.e., $\epsilon_a=1-1/K$, when the true label $y_0$ can be as likely as any other label, i.e., $p_0^*=1/K$ under the true distribution $p^*(x,y)$.

The aleatoric uncertainty imposes an irreducible upper bound on the accuracy of any classification model. When the aleatoric uncertainty is large, the selected input $x_0$ is inherently hard to classify, even when knowing the true distribution $p^*(x,y)$. Conversely, when the aleatoric uncertainty is small, the example is inherently easy to classify, at least when a sufficiently large training data set is available.

\vspace{-0.3cm}
\subsection{Calibration}
While the true probability of the class $y_0$ is $p_0^*$, the model assigns to it a probability $p_0$. A \textit{well-calibrated} model is expected to output a confidence probability $p_0$ that is close to the ground-truth probability $p_0^*$ \cite{guo2017calibration, cohen2023calibrating}. However, machine learning models tend to be \textit{overconfident}, assigning a probability $p_0$ larger than $p_0^*$, especially when trained with the example $(x_0,y_0)$ being evaluated. In fact, it is well known that the level of overconfidence of the target model is a significant factor influencing the power of MIA \cite{carlini2022membership,ye2022enhanced,shokri2017membership,chen2023overconfidence, salem2018ml}. A more overconfident model tends to assign a larger probability $p_0$, making it easier to detect the presence of example $(x_0,y_0)$ in the training data set.

The \textit{calibration error} is defined as the average difference between the estimated confidence probability and the ground-truth probability of the true label \cite{guo2017calibration, cohen2023calibrating}. In order to quantify the calibration performance of the target model, we define the \textit{relative calibration error} as
\begin{align}\label{RCE}
    \Delta = \frac{\mathbb{E}_{f^{\mathrm{in}}}\left[p_0\right]-p_0^*}{p_0^*},
\end{align}
which measures the relative difference between the expected confidence $\mathbb{E}_{f^\mathrm{in}}\left[p_0\right]$ generated by models trained over data sets that include the data point $(x_0,y_0)$ and the ground-truth probability $p^*_0\in[1/K,1]$.
As mentioned, an MIA is expected to be more likely to succeed when the model is overconfident, i.e., when the relative calibration error $\Delta$ is large.

\vspace{-0.3cm}
\section{Analysis of the Attacker's Advantage}\label{analysis}
In this section, we analyze the performance of LiRA in terms of the attacker's advantage \eqref{eq_Adv} for the three types of outputs defined in Sec. \ref{testing} (see also Fig. \ref{diff_observation}). We start by modeling the distributions $f^{\mathrm{out}}(O)$ and $f^{\mathrm{in}}(O)$ for the baseline case of CV observations in which the output $O$ equals the entire probability vector $p$ in \eqref{entire_v}.

\vspace{-0.3cm}
\subsection{Modeling the Output Distributions}\label{assumptions}
As explained in the previous section, the performance of LiRA hinges on the difference between the distributions $f^{\mathrm{out}}(O)$ and $f^{\mathrm{in}}(O)$ of the observations $O$ at the attacker, with $f^{\mathrm{out}}(O)$ denoting the observation distribution when the example $(x_0,y_0)$ is excluded from the model's training data set, while $f^{\mathrm{in}}(O)$ refers to the distribution when the target model is trained using the example $(x_0,y_0)$. Recall that randomness arises in LiRA due to the unknown data sets $\mathcal{D}^{\mathrm{out}}$ and $\mathcal{D}^{\mathrm{in}}$, which must be considered as random quantities as per \eqref{test}.

All output types \eqref{obs_p}--\eqref{obs_q_p} are functions of the confidence vector $p$. Therefore, the analysis of LiRA hinges on modeling the distributions $f^{\mathrm{out}}(p)$ and $f^{\mathrm{in}}(p)$. The \textit{Dirichlet distribution} provides a flexible probability density function for probability distributions over $K$ possible outputs \cite{kotz2004continuous}. Accordingly, while other choices are possible, we assume for our analysis that the distributions $f^{\mathrm{out}}(p)$ and $f^{\mathrm{in}}(p)$ are Dirichlet probability density functions with different parameter vectors $\gamma^{\mathrm{out}}=[\gamma_0^{\mathrm{out}},\ldots,\gamma_{K-1}^{\mathrm{out}}]^{\mathrm{T}}$ and $\gamma^{\mathrm{in}}=[\gamma_0^{\mathrm{in}},\ldots,\gamma_{K-1}^{\mathrm{in}}]^{\mathrm{T}}$, respectively. Sec. \ref{experiments} will provide a numerical example to illustrate this assumption. Therefore, the probability density functions of the confidence vectors output by the models trained on $\mathcal{D}^{\mathrm{out}}$ or $\mathcal{D}^{\mathrm{in}}$ are given by
\begin{equation}\label{Dirichlet_in_out}
    f^z(p)=\frac{1}{B(\gamma^z)}\prod_{k=0}^{K-1} p_k^{\gamma^z_k-1},~~\mathrm{for}~z\in\left\{\mathrm{out},\mathrm{in}\right\},
\end{equation}
where $B(\gamma^z)=\prod_{i=0}^{K-1}\Gamma(\gamma^z_i)\big/\Gamma(\sum_{i=0}^{K-1} \gamma^z_i)$ is the multivariate Beta function expressed in terms of the Gamma function $\Gamma(x)=\int_0^{+\infty}t^{x-1}e^{-t}dt$. Note that the theoretical analysis presented in this paper can be readily extended to account for alternative output distribution models, such as the logit-normal distribution \cite{atchison1980logistic}.

Parameters $\gamma_0^{\mathrm{out}}$ and $\gamma_0^{\mathrm{in}}$ control the \textit{average confidence} assigned to the true label by the models trained without and with $(x_0,y_0)$, while the remaining parameters $\gamma^{\mathrm{out}}_1,\ldots,\gamma^{\mathrm{out}}_{K-1}$ and $\gamma^{\mathrm{in}}_1,\ldots,\gamma^{\mathrm{in}}_{K-1}$ reflect the average confidence levels assigned by the two models to the remaining $K-1$ candidate labels. More precisely, the \textit{average} confidence levels are given by \cite{kotz2004continuous}
\begin{equation}\label{average}
    \mathbb{E}_{f^z}\left[p_k\right]=\frac{\gamma^z_k}{\sum_{i=0}^{K-1}\gamma^z_i}~\mathrm{for}~k=0,1,\ldots, K-1,
\end{equation}
where $z\in\left\{\mathrm{out},\mathrm{in}\right\}$. Considering that the confidence probability of the true label constitutes the primary source of information for the attacker, we set $\gamma^{\mathrm{out}}_1=\cdots=\gamma^{\mathrm{out}}_{K-1}$ and $\gamma^{\mathrm{in}}_1=\cdots=\gamma^{\mathrm{in}}_{K-1}$ for all other candidate classes. This assumption will facilitate the interpretation of the results of the analysis, although it is not necessary for our derivations.

Being random due to the ignorance of the attacker about the training data set, the confidence levels $p_k$ have variability that can be measured by the variance of the Dirichlet distribution
\begin{align}\label{variance}
    \mathrm{Var}_{f^z}[p_k] = \frac{\gamma_k^z\big(\sum_{i=0}^{K-1}\gamma^z_i - \gamma_k^z\big)}{\big(\sum_{i=0}^{K-1}\gamma^z_i\big)^2\big(\sum_{i=0}^{K-1}\gamma^z_i + 1\big)}&\nonumber\\
    \mathrm{for}~k=0,1,\ldots,K-1,&
\end{align}
with $z\in\left\{\mathrm{out}, \mathrm{in}\right\}$. The variance \eqref{variance} accounts for the dependence of the trained model as a function of the training sets drawn from the ground-truth distribution $p^*(x,y)$. This model-level uncertainty, known as \textit{epistemic uncertainty} \cite{hullermeier2021aleatoric}, decreases as the size $N^{\mathrm{tr}}$ of the training data set increases and as the model's capacity to learn the underlying data distribution improves. In fact, in the regime $N^{\mathrm{tr}}\rightarrow \infty$ where the training data sets are fully representative of the distribution $p^*(x,y)$, and assuming the model has sufficient capacity, the learned probabilities $p_k$ tend to the ground-truth values $p^*_k$ \cite{bishop2006pattern, goodfellow2016deep}. With a finite training set, an attacker following Definition \ref{def_MIG} is characterized by uncertainty on the model, which we account for in our analysis through the variance of the model output distribution \cite{valdenegro2022deeper}.

The possibility to decrease the epistemic uncertainty is in stark contrast to the aleatoric uncertainty introduced in the previous section, which is due to the inherent randomness of the data, and thus it does not decrease as the training set size or the model capacity increases. It is therefore possible to simultaneously have a high epistemic uncertainty due to limited data and a low aleatoric uncertainty for ``easy'' examples; and, vice versa, a high aleatoric uncertainty for ``hard'' examples and a low epistemic uncertainty due to access to a large training set.

To account for epistemic uncertainty, we adopt the reciprocal of the sum of Dirichlet distribution parameters, which is proportional to variance \eqref{variance}, i.e.,
\begin{equation}\label{EU}
    \epsilon_e = \frac{1}{\sum_{i=0}^{K-1}\gamma_i^{\mathrm{out}}}= \frac{1}{\sum_{i=0}^{K-1}\gamma_i^{\mathrm{in}}}.
\end{equation}
Furthermore, by \eqref{EU}, since both the data set size $N^{\mathrm{tr}}$ and the model class $\mathcal{F}$ in \eqref{eq_mod_cla} are the same for both data sets $\mathcal{D}^{\mathrm{out}}$ and $\mathcal{D}^{\mathrm{in}}$, we assume the sum in \eqref{EU} to be the same for both distributions $f^{\mathrm{out}}(p)$ and $f^{\mathrm{in}}(p)$.

Finally, using \eqref{AU}--\eqref{EU}, we can write the Dirichlet parameters as a function of the aleatoric uncertainty $\epsilon_a$ in \eqref{AU} and epistemic uncertainty $\epsilon_e$ in \eqref{EU} as
\begin{subequations}\label{gamma_out}
    \begin{align}
        &\gamma^{\mathrm{out}}_0 = \frac{1-\epsilon_a}{\epsilon_e}, \label{gamma_out_0}\\
        \mathrm{and}~&\gamma^{\mathrm{out}}_k = \frac{\epsilon_a}{(K-1)\epsilon_e}~\mathrm{for}~k=1,2,\ldots,K-1,\label{gamma_out_k}
    \end{align}
\end{subequations}
for distribution $f^{\mathrm{out}}(p)$; while, for distribution $f^{\mathrm{in}}(p)$, accounting also for the relative calibration error \eqref{RCE}, we have
\begin{subequations}\label{gamma_in}
    \begin{align}
        &\gamma^{\mathrm{in}}_0 = \frac{(1+\Delta)(1-\epsilon_a)}{\epsilon_e}, \label{gamma_in_0}\\
        \mathrm{and}~&\gamma^{\mathrm{in}}_k = \frac{\epsilon_a\Delta+\epsilon_a-\Delta}{(K-1)\epsilon_e}~\mathrm{for}~k=1,2,\ldots,K-1.\label{gamma_in_k}
    \end{align}
\end{subequations}
Please refer to Appendix \ref{Appendix_eq} for a derivation of \eqref{gamma_out} and \eqref{gamma_in}. Note that, in order to ensure the inequalities $\gamma^z_k>0$ with $z\in\left\{\mathrm{out},\mathrm{in}\right\}$ and $k =0,1,\ldots,K-1$, we have the inequality $\Delta/(1+\Delta)<\epsilon_a<1$. Equations \eqref{gamma_out}--\eqref{gamma_in} indicate that the average confidence generated by the model in the true label $y_0$ decreases with the aleatoric uncertainty $\epsilon_a$ and with epistemic uncertainty $\epsilon_e$, while also increasing with the relative calibration error $\Delta$ for the model trained with $(x_0,y_0)$.

\vspace{-0.3cm}
\subsection{Confidence Vector Disclosure}\label{analysis_CV}
In this subsection, we leverage Lemma \ref{lemma_ub} to obtain an upper bound on the advantage of the attacker in the worst case for the target model in which the attacker observes the entire CV, i.e., $O(p)=p$. We have the following result.

\begin{proposition}\label{Prop_1}
    The advantage of the attacker with CV disclosure \eqref{obs_p} can be upper bounded as
    \begin{align}\label{CV_UB}
        \mathrm{Adv}^{\mathrm{\rm{CV}}}_{\alpha} \leq& \Big(\frac{\Delta(1-\epsilon_a)}{\epsilon_e}\Big(\psi\Big(\frac{(1+\Delta)(1-\epsilon_a)}{\epsilon_e}\Big)-\psi\Big(\frac{1-\epsilon_a}{\epsilon_e}\Big)\nonumber\\
        &+ \psi\Big(\frac{\epsilon_a}{(K-1)\epsilon_e}\Big)-\psi\Big(\frac{\epsilon_a(1+\Delta)-\Delta}{(K-1)\epsilon_e}\Big)\Big)\Big)^{\frac{1}{2}} \nonumber\\
        \triangleq& \mathrm{Adv}_{\alpha}^{\mathrm{CV-ub}},
    \end{align}
    where $\psi(x)=\Gamma'(x)/\Gamma(x)$ denotes the digamma function. Furthermore, using the approximation $\psi(x)\approx \ln x- 1/2x$, which holds when $x\rightarrow\infty$, we obtain the approximate upper bound
    \begin{align}\label{approx_ub}
        \mathrm{Adv}_{\alpha}^{\mathrm{\rm{CV-ub}}} \approx& \Big(\frac{\Delta(1-\epsilon_a)}{\epsilon_e}\ln\Big(\frac{(1+\Delta)\epsilon_a}{(1+\Delta)\epsilon_a-\Delta}\Big) \nonumber\\
        &+\frac{\Delta^2(1-\epsilon_a)^2(K-1)}{2\epsilon_a((1+\Delta)\epsilon_a-\Delta)}+\frac{\Delta^2}{2(1+\Delta)}\Big)^{\frac{1}{2}}\nonumber\\
        \triangleq& \mathrm{A\widetilde{d}v}_{\alpha}^{\mathrm{\rm{CV-ub}}}.
    \end{align}
\end{proposition}
\textit{Proof:} The proof of \eqref{CV_UB} is detailed in Appendix \ref{Appendix_prop1}.

The approximate bound in \eqref{approx_ub} can be used to draw some analytical insights into the attacker's advantage, which will be validated in Sec. \ref{experiments} by experiments. First, the bound \eqref{approx_ub} increases as the relative calibration error $\Delta$ grows, which is in line with empirical evidence that overconfident models are more vulnerable to MIA \cite{carlini2022membership, shokri2017membership, ye2022enhanced, chen2023overconfidence}.

Second, the bound \eqref{approx_ub} decreases as the aleatoric uncertainty $\epsilon_a$ and/or the epistemic uncertainty $\epsilon_e$ increase. Both a larger aleatoric uncertainty and a larger epistemic uncertainty tend to yield less confident, higher-entropy predictions, which in turn mitigates model overconfidence \cite{hullermeier2021aleatoric}. This ensures that the model reveals less information about the training data, potentially reducing the attack's effectiveness. This conclusion aligns with the experimental findings in \cite{becker2022evaluating}, where the epistemic uncertainty was increased through unlearning in order to reduce the model's information about the target data, thereby potentially reducing the effectiveness of MIA.

\vspace{-0.3cm}
\subsection{True Label Confidence Disclosure}\label{analysis_TLC}
Consider now the case in which the attacker has only access to the confidence probability corresponding to the true label $y_0$ given input $x_0$, i.e., $O(p)=p_0$, as in \cite{carlini2022membership}. The marginal probability density function for variable $p_0$ from the joint Dirichlet distribution \eqref{Dirichlet_in_out} follows the Beta distribution. Accordingly, the probability density functions of the TLC observations output by models trained on data sets that exclude and include, respectively, the target point $(x_0,y_0)$, are given by
\begin{equation}\label{Beta_in_out}
    f^z(p_0;\gamma^z_0,\overline{\gamma}^z_0)=\frac{1}{B(\gamma^z_0, \overline{\gamma}^z_0)}p_0^{\gamma^z_0}(1-p_0)^{\overline{\gamma}^z_0-1}~\mathrm{for}~z\in\left\{\mathrm{out},\mathrm{in}\right\},
\end{equation}
with $\overline{\gamma}^z_0=\sum_{k=1}^{K-1}\gamma^z_k$, Beta function $B(\gamma^z_0, \overline{\gamma}^z_0) = \Gamma(\gamma^z_0) \Gamma(\overline{\gamma}^z_0) \big/\Gamma(\gamma^z_0+\overline{\gamma}^z_0)$, and Gamma function $\Gamma(x)=\int_0^{+\infty}t^{x-1}e^{-t}dt$. This observation yields the following result.

\begin{proposition}\label{Prop_2}
    The advantage of the attacker with TLC disclosure \eqref{obs_p0} can be upper bounded as
    \begin{align}\label{TLC_ub}
        \mathrm{Adv}^{\mathrm{\rm{TLC}}}_{\alpha} \leq& \Big(\frac{\Delta(1-\epsilon_a)}{\epsilon_e}\Big(\psi\Big(\frac{(1+\Delta)(1-\epsilon_a)}{\epsilon_e}\Big)-\psi\Big(\frac{1-\epsilon_a}{\epsilon_e}\Big)\nonumber\\
        &+ \psi\Big(\frac{\epsilon_a}{\epsilon_e}\Big)-\psi\Big(\frac{\epsilon_a(1+\Delta)-\Delta}{\epsilon_e}\Big)\Big)\Big)^{\frac{1}{2}}\nonumber\\
        \triangleq& \mathrm{Adv}^\mathrm{TLC-ub}_{\alpha},
    \end{align}
    where $\psi(x)=\Gamma'(x)/\Gamma(x)$ denotes the digamma function. Furthermore, using the approximation $\psi(x)\approx \ln x- 1/2x$, which holds when $x\rightarrow\infty$, we obtain the approximate upper bound
    \begin{align}\label{approx_ub_TLC}
         \mathrm{Adv}^\mathrm{TLC-ub}_{\alpha} \approx& \Big(\frac{\Delta(1-\epsilon_a)}{\epsilon_e}\ln\Big(\frac{(1+\Delta)\epsilon_a}{(1+\Delta)\epsilon_a-\Delta}\Big)\nonumber\\
         &+\frac{\Delta^2(1-\epsilon_a)^2}{2\epsilon_a((1+\Delta)\epsilon_a-\Delta)}+\frac{\Delta^2}{2(1+\Delta)}\Big)^{\frac{1}{2}}\nonumber\\
         \triangleq&\mathrm{A\widetilde{d}v}^\mathrm{TLC-ub}_{\alpha}.
    \end{align}
\end{proposition}
\textit{Proof:} The upper bound $\mathrm{Adv}_{\alpha}^{\mathrm{TLC-ub}}$ in \eqref{TLC_ub} and the approximate upper bound $\mathrm{A\widetilde{d}v}_{\alpha}^{\mathrm{TLC-ub}}$ in \eqref{approx_ub_TLC} can be obtained by substituting $K=2$ into \eqref{CV_UB} and \eqref{approx_ub} in Proposition \ref{Prop_1}, respectively.

By the bound \eqref{approx_ub_TLC}, the MIA advantage with TLC disclosure exhibits the same general trends with respect to calibration error $\Delta$ and uncertainties $\epsilon_a$ and $\epsilon_e$ as for CV disclosure. Furthermore, comparing \eqref{approx_ub_TLC} with \eqref{approx_ub}, the MIA advantage under CV observations is larger than that under TLC observations according to the derived bounds, with the gap between the two bounds growing as the number of classes, $K$, increases.

\vspace{-0.3cm}
\subsection{Decision Set Disclosure}\label{analysis_DS}
In this subsection, we study the DS disclosure scenarios. To this end, we analyze a generalization of \eqref{obs_q_p} that allows for randomization. Randomization is a well-established mechanism to ensure privacy \cite{youn2023randomized, dwork2014algorithmic, liu2020privacy}, and thus it is interesting to investigate its potential role in MIA.

\begin{figure}[t]
    \centering
    \setlength{\abovecaptionskip}{-2pt}
    {\includegraphics[width = 0.32\textwidth]{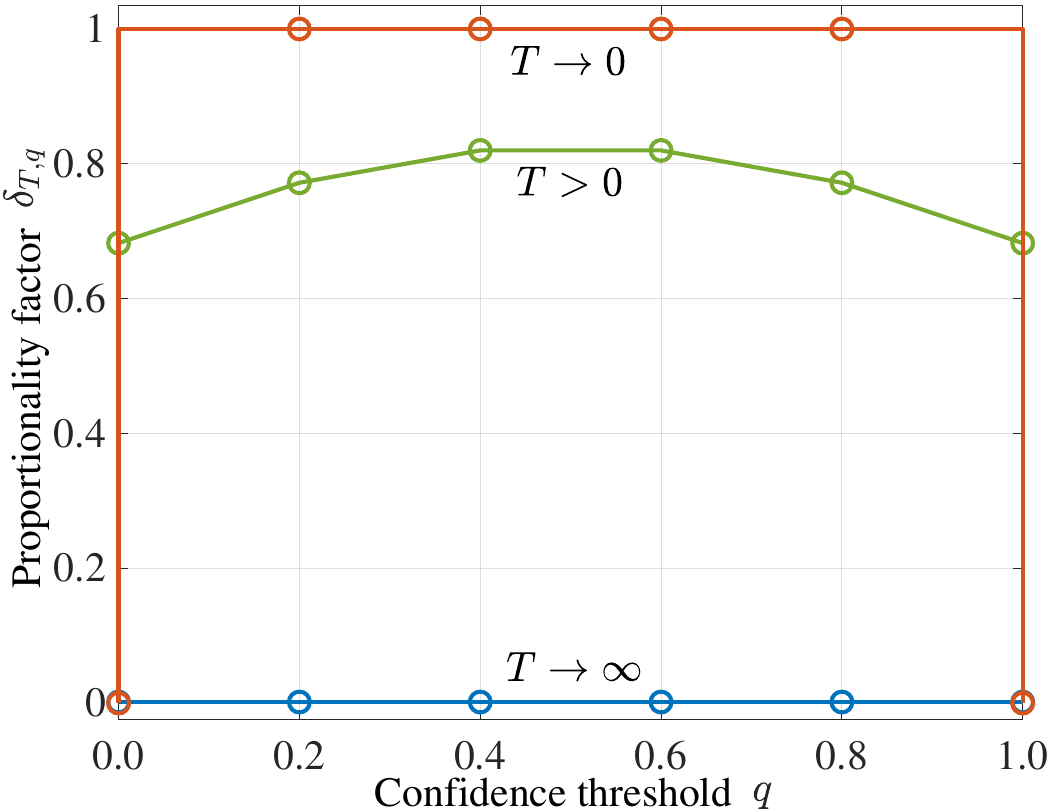}}
    \caption{The proportionality factor $\delta_{T,q}$ between the advantage of DS disclosure \eqref{obs_p_q_random} and CV disclosure \eqref{obs_p} as a function of the confidence threshold $q$ for different values of the temperature parameter $T$ with $K=2$.
    }\label{TV_thre_fig}
    \vspace{-5mm}
\end{figure}

To this end, we consider the observation
\begin{equation}\label{obs_p_q_random}
    O(p)=
    \begin{bmatrix}
        b_0 \\
        \vdots \\
        b_{K-1}
    \end{bmatrix}
    =
    \mathds{1}_T(p\geq q)
\end{equation}
where $\mathds{1}_T(p\geq q)$ represents a stochastic element-wise threshold function with temperature parameter $T>0$. This function returns random variables $b_k\in\{0,1\}$ with
\begin{equation}\label{p_thre}
    b_k =
    \begin{cases}
        1 & \text{with probability $\sigma_T(p_k-q)$},\\
        0 & \text{with probability $\sigma_T(q-p_k)$},
    \end{cases}
\end{equation}
where $\sigma_T(x)=1/(1+e^{-x/T})$ represents the sigmoid function with a temperature parameter $T>0$ \cite{papernot2021tempered} and $b_k=1$ indicates the inclusion of the $k$-th label in the decision set. The stochastic DS \eqref{obs_p_q_random} reduces to \eqref{obs_q_p} in the limit $T\rightarrow 0$.
\begin{proposition}\label{Prop_3}
    The advantage of the attacker with DS disclosure \eqref{obs_p_q_random} can be upper bounded as
    \begin{align}\label{DS_UB}
        \mathrm{Adv}^{\mathrm{\rm{DS}}}_{\alpha} \leq \delta_{T,q}\cdot\mathrm{Adv}^{\mathrm{CV-ub}}_{\alpha}& \triangleq \mathrm{Adv}^{\mathrm{DS-ub}}_{\alpha},
    \end{align}
    where 
    \begin{align}\label{TV_defi}
        \delta_{T,q}=&\Big(\max_{p,p'}\frac{1}{2}\sum_{b\in\{0,1\}^K}\Big|\prod_{k=0}^{K-1}\sigma_T\big((2b_k-1)(p_k-q)\big)\nonumber\\
        &-\prod_{k=0}^{K-1}\sigma_T\big((2b_k-1)(p'_k-q)\big)\Big|\Big)^{\frac{1}{2}},
    \end{align}
    with maximization over $k\times 1$ probability vectors $p$ and $p'$, and $\mathrm{Adv}^{\mathrm{CV-ub}}_{\alpha}$ is defined in \eqref{CV_UB}. Using \eqref{approx_ub} in Proposition \ref{Prop_1}, we can then obtain the approximate upper bound
    \begin{align}\label{approx_ub_DS}
        \mathrm{A\widetilde{d}v}^{\mathrm{DS-ub}}_{\alpha} \approx \delta_{T,q}\cdot \mathrm{A\widetilde{d}v}^{\mathrm{CV-ub}}_{\alpha},
    \end{align}
    where $\mathrm{A\widetilde{d}v}^{\mathrm{CV-ub}}_{\alpha}$ is defined in \eqref{approx_ub}.
\end{proposition}
\textit{Proof:} The proof of \eqref{DS_UB} is detailed in Appendix \ref{Appendix_prop3}.

While the upper bound \eqref{approx_ub_DS} requires numerical optimization to evaluate \eqref{TV_defi}, some useful insights can be obtained from \eqref{approx_ub_DS}. In particular, the bound indicates that the attacker's advantage under DS disclosure is proportional to that under the worst-case (for the target model) CV disclosure with a proportionality constant $\delta_{T,q}$. This constant can be proved to satisfy the inequality
\begin{align}
    \delta_{T,q}\leq 1,
\end{align}
reflecting the weaker power of the adversary under DS disclosure.

\begin{figure*}[t]
    \centering
    \setlength{\abovecaptionskip}{-2pt}
    {
    \includegraphics[width = 0.32\textwidth]{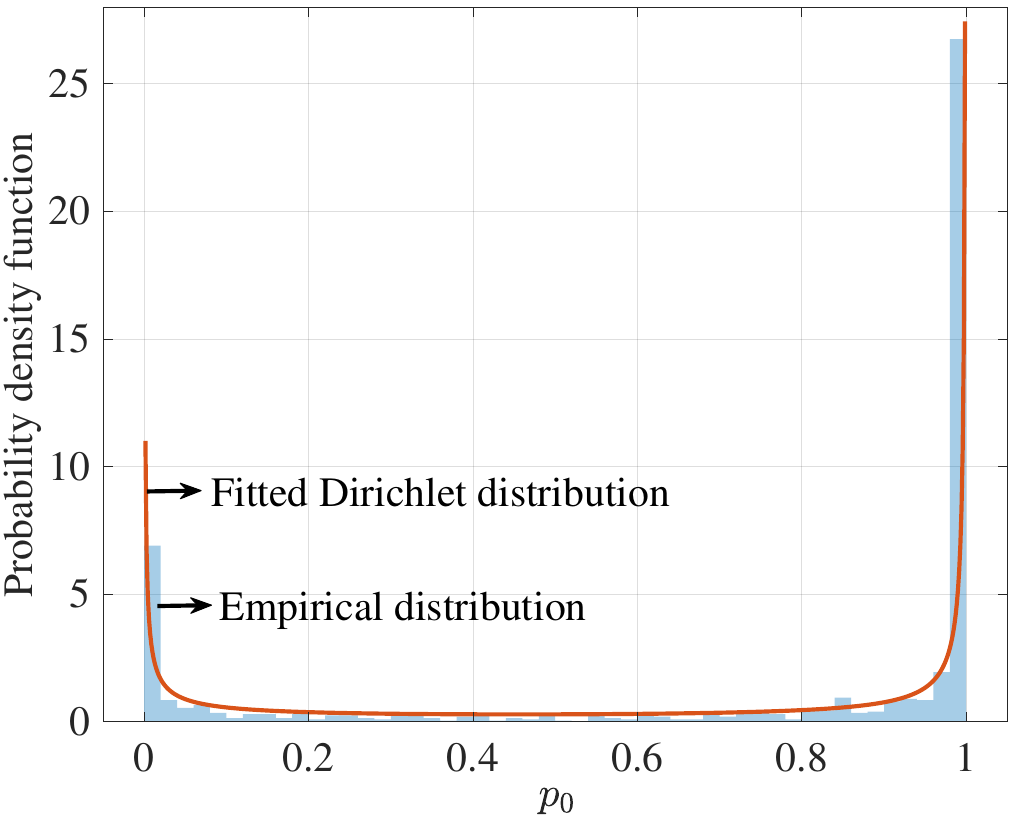}
	\includegraphics[width = 0.32\textwidth]{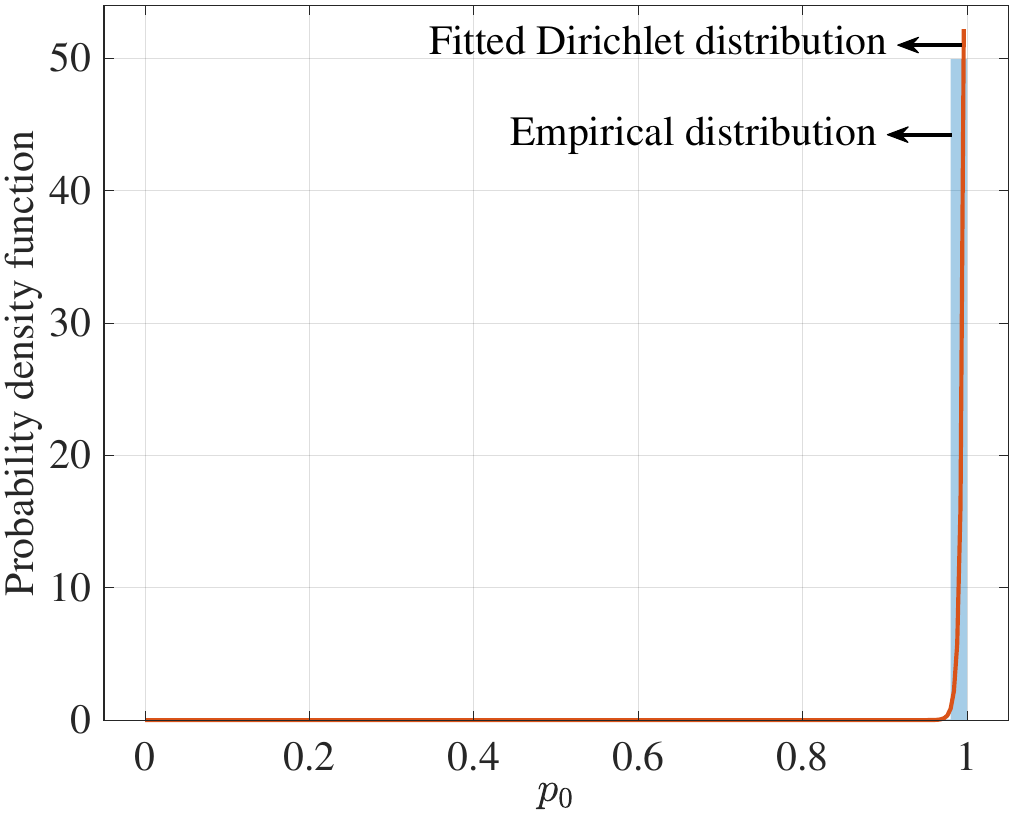}
	}
    \caption{Empirical distributions of the confidence level $p_0$ and corresponding marginal of the fitted Dirichlet distributions for models trained without (left) and with (right) the target sample $(x_0,y_0)$.}\label{fitting}
    \vspace{-5mm}
\end{figure*}
\begin{figure*}[t]
    \centering
    \setlength{\abovecaptionskip}{-2pt}
    {
	\includegraphics[width = 0.32\textwidth]{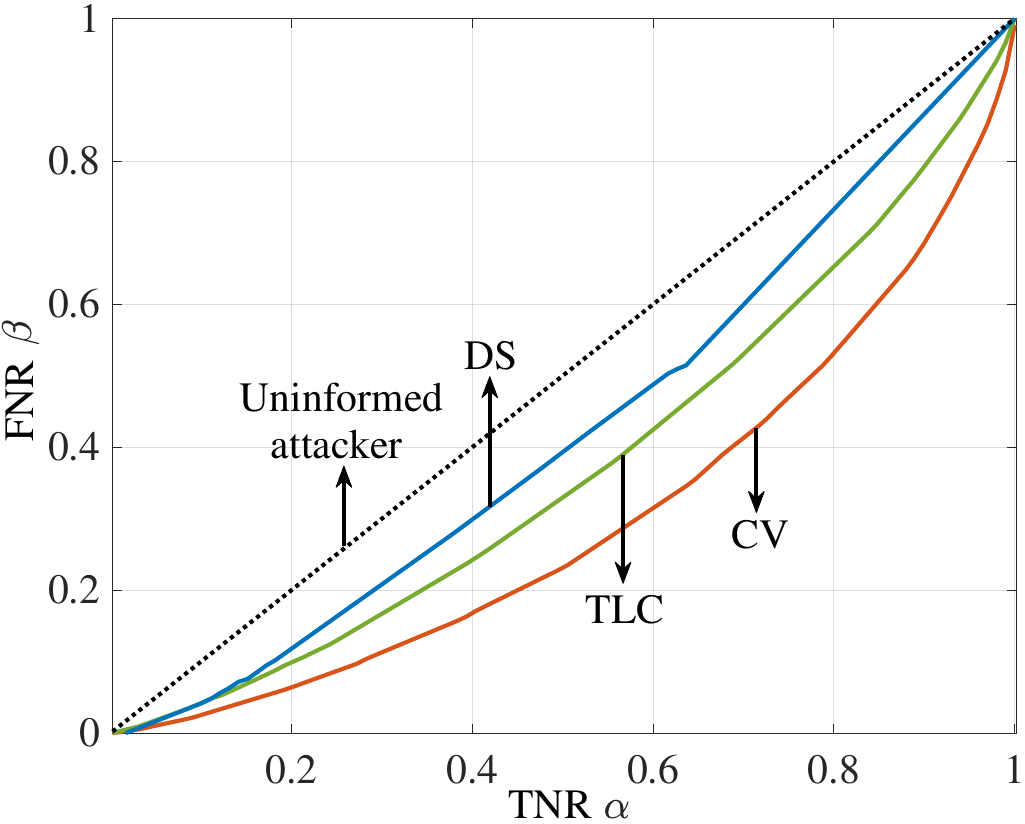}
	\includegraphics[width = 0.32\textwidth]{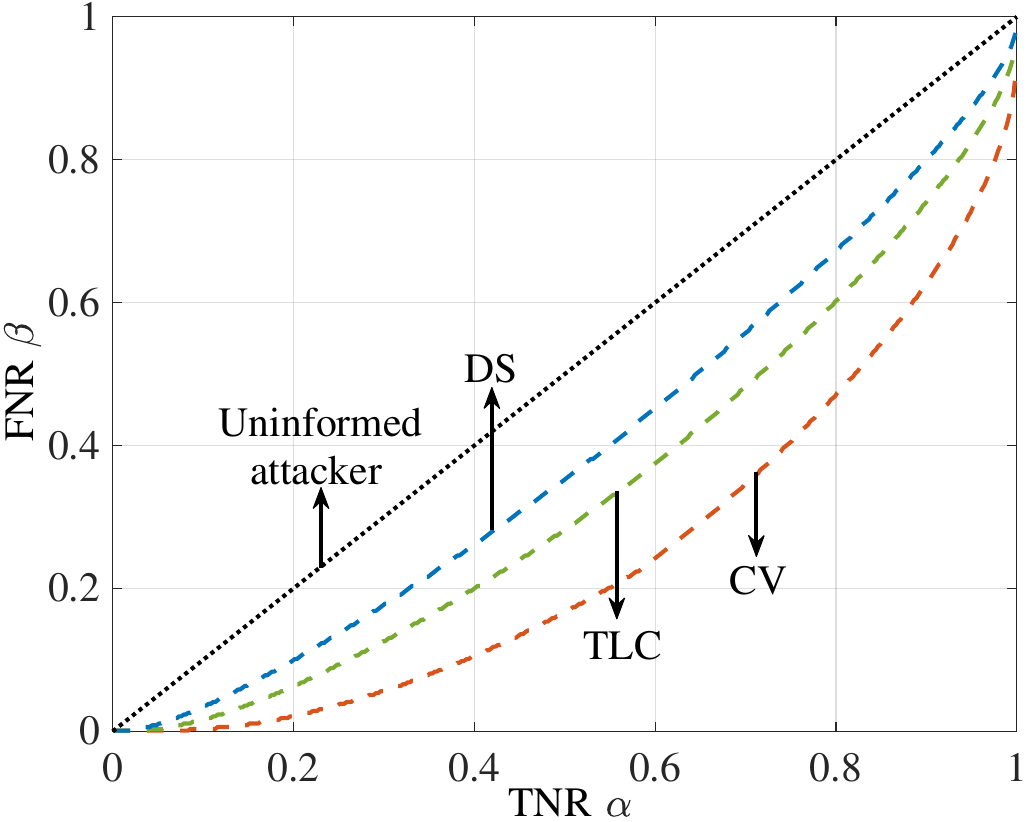}
    }
    \caption{True trade-off function $\beta_{\alpha}$ in \eqref{beta_LRT} (left) and the lower bound $\beta^{\mathrm{lb}}_{\alpha}$ in \eqref{lower_bound1} (right) for CV, TLC, and DS observations with relative calibration error $\Delta=0.2$, aleatoric uncertainty $\epsilon_a=0.5$, epistemic uncertainty $\epsilon_e=0.25$, and threshold $q = 0.2$.}\label{ROC_NP_KL}
    \vspace{-5mm}
\end{figure*}
The proportionality factor $\delta_{T,q}$ depends on the temperature parameter $T$ as illustrated in Fig. \ref{TV_thre_fig} for the case $K=2$. In particular, for $T\rightarrow 0$, which corresponds to the deterministic function in \eqref{obs_q_p}, the constant attains the maximum value of $1$, except for the trivial cases $q=0$ and $q=1$, which yield $\delta_{T,q}=0$ as $T\rightarrow 0$. Thus, for a deterministic disclosure function, the upper bound \eqref{approx_ub_DS} is not sufficiently tight to quantify any decrease in the attacker's advantage. However, for any $T>0$, and thus even with minimal randomness, the bound predicts that the attacker's advantage at first increases as $q$ grows, since a larger threshold in \eqref{obs_p_q_random} allows to capture more information about the confidence vector $p$. However, an excessively large $q$ ends up decreasing the performance of the attacker as the predicted set $O(p)$ in \eqref{obs_p_q_random} gets increasingly smaller. Furthermore, as $T\rightarrow \infty$, the sigmoid function equals $\sigma_T(x)=1/2$ and the proportionality factor equals zero, $\delta_{T,q}=0$. In this case, the attacker does not obtain any useful information, resulting in a vanishing advantage.

\vspace{-3mm}
\section{Numerical Results} \label{experiments}
In this section, we provide numerical results to validate the analysis. We begin by illustrating the assumption of modeling the confidence probability vector using a Dirichlet distribution. Subsequently, we compare the analytical bounds to the actual performance of attacks based on the observations \eqref{obs_p}--\eqref{obs_q_p} as a function of calibration and uncertainty metrics. Furthermore, for attacks leveraging DS observation, we further evaluate the trade-off between prediction set size, which determines the informativeness of the predictor \cite{zecchin2024generalization} and the attacker's advantage. Throughout, we consider the target model to be a classifier with $K=10$ classes.

\vspace{-0.3cm}
\subsection{On the Dirichlet Assumption}\label{subsec_Dir}
First, to validate the use of a Dirichlet distribution to model the confidence probability, we evaluate the true distribution of the confidence vector produced by a neural network-based classifier. Following the setting in \cite{shokri2017membership, carlini2022membership, ali2023membership, ye2022enhanced, zarifzadeh2023low}, we use the CIFAR-10 data set, a standard benchmark for image classification involving images classified by $K=10$ labels. To construct the training sets $\mathcal{D}^{\mathrm{out}}$ and $\mathcal{D}^{\mathrm{in}}$, we draw $N^{\mathrm{tr}}=4000$ samples at random from the overall data set. The target classification model adopts a standard convolutional neural network (CNN) with two convolution and max pooling layers, followed by a $128$-unit fully connected layer and a SoftMax output layer. The Tanh is used as the activation function, and the cross-entropy serves as the loss function. We adopt the stochastic gradient descent (SGD) optimizer with a learning rate of $0.001$, momentum of $0.9$, and weight decay of $1\times10^{-7}$, over $100$ epochs for model training.

We collect probability vectors generated by the trained model for images corresponding to the training and test images with the truth label $0$. We then apply maximum likelihood estimation with the L-BFGS-B algorithm, a gradient descent variant tailored for large-scale optimizations \cite{byrd1995limited}, to fit the generated data points to Dirichlet distributions for both models trained on $\mathcal{D}^{\mathrm{out}}$ and $\mathcal{D}^{\mathrm{in}}$.

Fig. \ref{fitting} illustrates the probability density functions of the confidence level $p_0$ for the target sample $(x_0,y_0)$, along with the corresponding empirical distribution obtained from the fitted Dirichlet distributions for the models trained on $\mathcal{D}^{\mathrm{out}}$ (left) and $\mathcal{D}^{\mathrm{in}}$ (right). The model is observed to be overconfident when trained by including the sample $(x_0,y_0)$. The empirical distributions closely align with the fitted Dirichlet distributions, confirming the flexibility of the working assumption of Dirichlet distributions for the confidence vectors. Throughout the rest of this section, we adopt this assumption, which is further validated in Appendix \ref{Appendix_fitting}.

\vspace{-0.3cm}
\subsection{On the Impact of Calibration and Uncertainty}\label{subsec_CE_AU_EU}
In this subsection, to validate the analysis in Sec. \ref{analysis}, we compare the true trade-off function $\beta_{\alpha}$ in \eqref{beta_LRT} along with the lower bound $\beta_{\alpha}^{\mathrm{lb}}$ in \eqref{lower_bound1} obtained by Lemma \ref{lemma_outer_bound} by setting the relative calibration error as $\Delta=0.2$, the aleatoric uncertainty as $\epsilon_a=0.5$, the and the epistemic uncertainty as $\epsilon_e=0.25$ in the Dirichlet parameters \eqref{gamma_out}--\eqref{gamma_in}. Fig. \ref{ROC_NP_KL} plots the true trade-off function (left) and lower bound (right) for all disclosure schemes. In particular, for DS, we consider the situation most vulnerable to MIA by applying a deterministic thresholding scheme, i.e., $T\rightarrow 0$ in \eqref{p_thre}.

\begin{figure}[t]
    \centering
    \setlength{\abovecaptionskip}{-2pt}
    {\includegraphics[width = 0.32\textwidth]{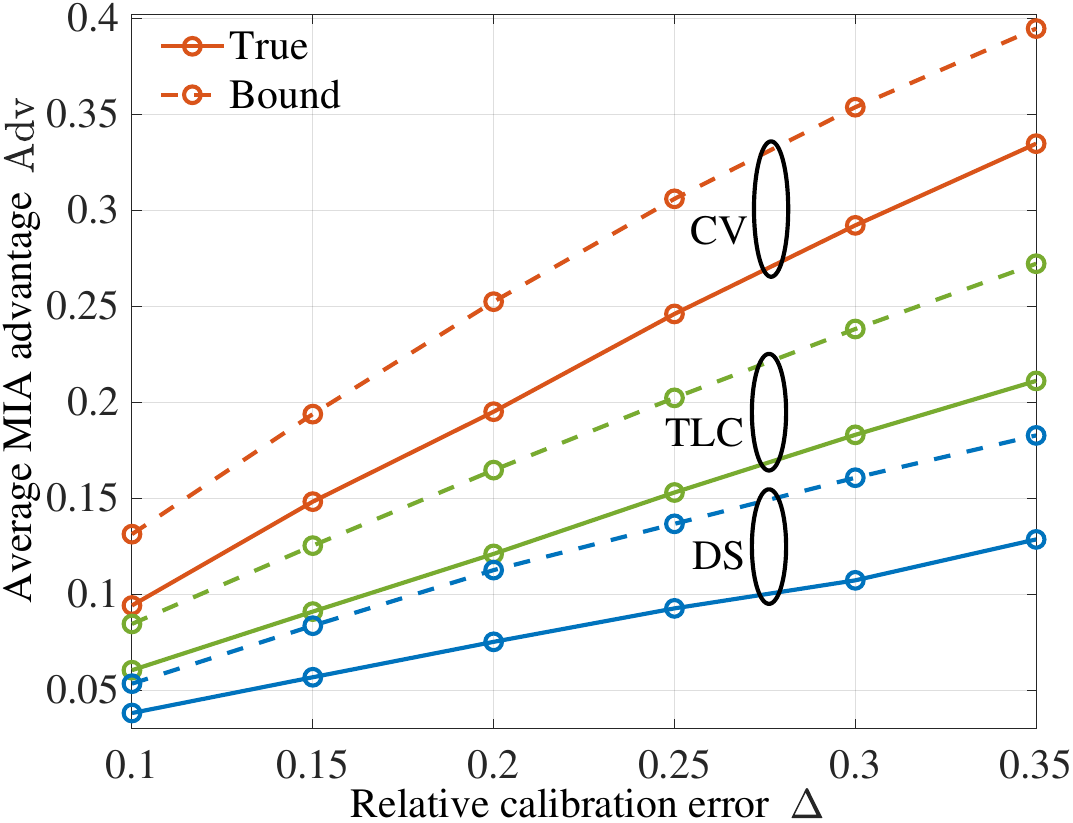}}
    \caption{True advantage \eqref{eq_Adv} and bounds \eqref{lower_bound2} for the attacker's advantage with CV, TLC, and DS observations as a function of relative calibration error $\Delta$ with aleatoric uncertainty $\epsilon_a=0.5$, epistemic uncertainty $\epsilon_e=0.25$, and threshold $q = 0.2$.}\label{Fig_Adv_RCE}
    \vspace{-5mm}
\end{figure}
\begin{figure}[t]
    \centering
    \setlength{\abovecaptionskip}{-2pt}
    {\includegraphics[width = 0.32\textwidth]{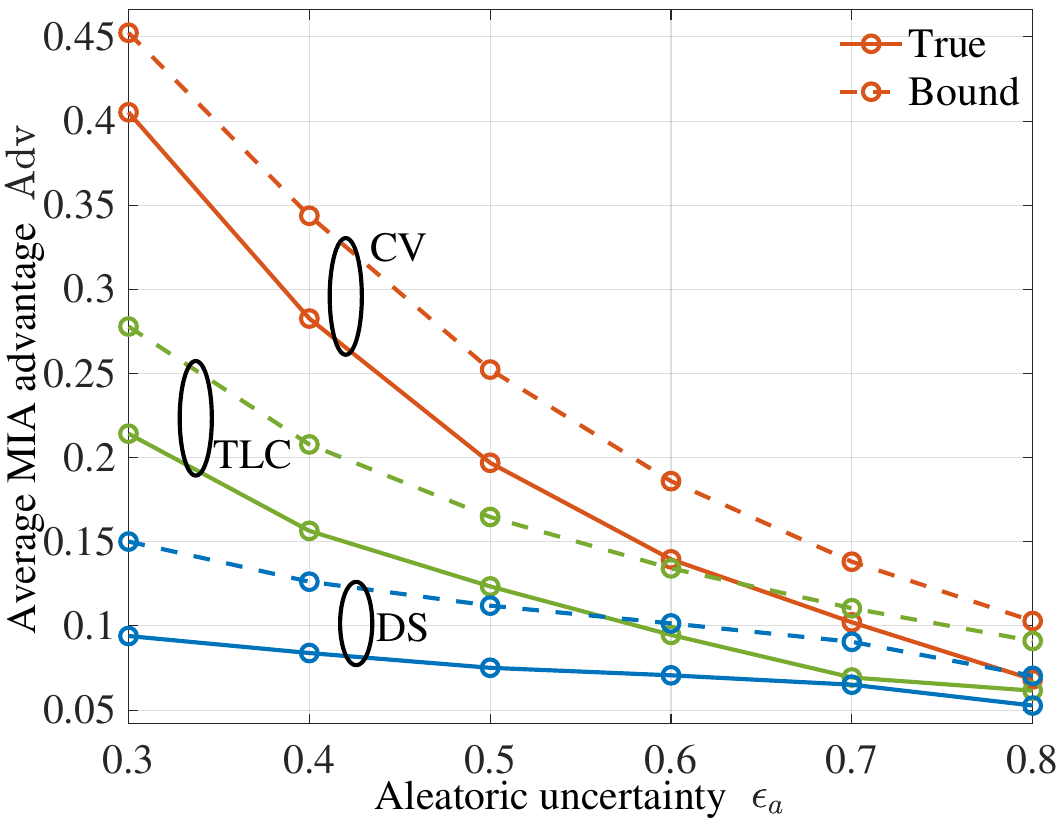}}
    \caption{True advantage \eqref{eq_Adv} and bounds \eqref{lower_bound2} for the attacker's advantage with CV, TLC, and DS observations as a function of aleatoric uncertainty $\epsilon_a$ with relative calibration error $\Delta=0.2$, epistemic uncertainty $\epsilon_e=0.25$, and threshold $q = 0.2$.}\label{Fig_Adv_AU}
    \vspace{-5mm}
\end{figure}
\begin{figure}[t]
    \centering
    \setlength{\abovecaptionskip}{-2pt}
    {\includegraphics[width = 0.32\textwidth]{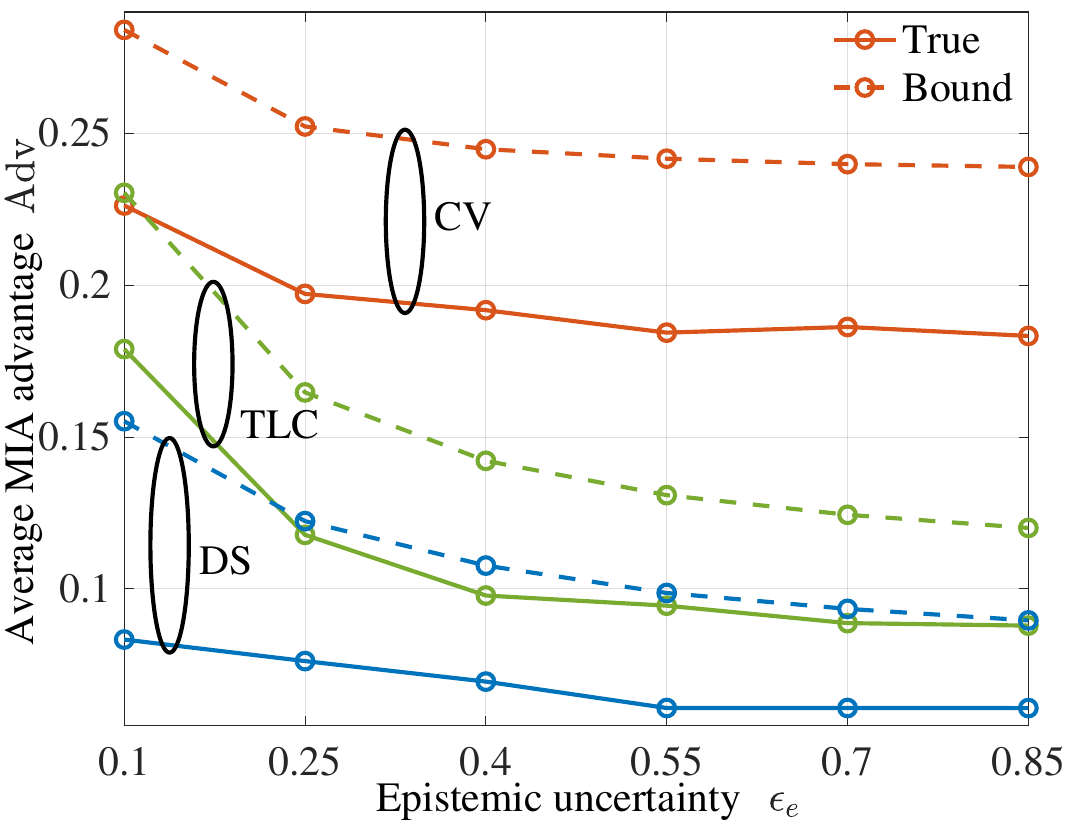}}
    \caption{True advantage \eqref{eq_Adv} and bounds \eqref{lower_bound2} for the attacker's advantage with CV, TLC, and DS observations as a function of epistemic uncertainty $\epsilon_e$ with relative calibration error $\Delta=0.2$, aleatoric uncertainty $\epsilon_a=0.5$, and threshold $q = 0.4$.}\label{Fig_Adv_EU}
    \vspace{-5mm}
\end{figure}

The figures confirm that the derived lower bound provides a useful prediction of the relative performance of the three schemes. Specifically, both the true attacker's advantage and the lower bound indicate that CV disclosure provides the most information to the attacker, followed by TLC and DS.

We now turn to studying the impact of calibration and uncertainty on the attacker's advantage. To this end, in Fig. \ref{Fig_Adv_RCE}, we first display the attacker's advantage averaged over a uniformly selected TNR $\alpha$ as $\mathrm{Adv} = \mathbb{E}_{\alpha\sim U(0,1)}[\alpha - \beta_{\alpha}]$, as a function of the relative calibration error $\Delta$ with $\epsilon_a=0.5$ and $\epsilon_e=0.25$. Bound \eqref{lower_bound2} in Lemma \ref{lemma_ub} and true values are again observed to be well aligned. Furthermore, confirming the empirical observations in \cite{carlini2022membership, shokri2017membership, ye2022enhanced, chen2023overconfidence}, the attacker's advantage is seen to increase with $\Delta$. In fact, overconfident models tend to output higher confidence probabilities for the true label when trained with the target sample, which can be easily distinguished from the confidence probabilities for the other labels. Irrespective of the relative calibration error, the attacker's advantage decreases as the model discloses less information via TLC and DS.

Fig. \ref{Fig_Adv_AU} depicts the average attacker's advantage with CV, TLC, and DS disclosures as a function of the aleatoric uncertainty $\epsilon_a$ with $\Delta = 0.2$ and $\epsilon_e=0.25$. A larger aleatoric uncertainty reflects greater uncertainty in the underlying data. Accordingly, the effectiveness of attacks based on all observation types decreases and gradually approaches the theoretical lower limit of $0$. This is because the distributions of the observations produced by models trained without or with the target sample tend to become closer as the inherent data uncertainty grows. This, in turn, reduces the attacker's ability to exploit the overconfidence in the target model for successful attacks.

\begin{figure*}[t]
    \centering
    \setlength{\abovecaptionskip}{-2pt}
    {\includegraphics[width = 0.32\textwidth]{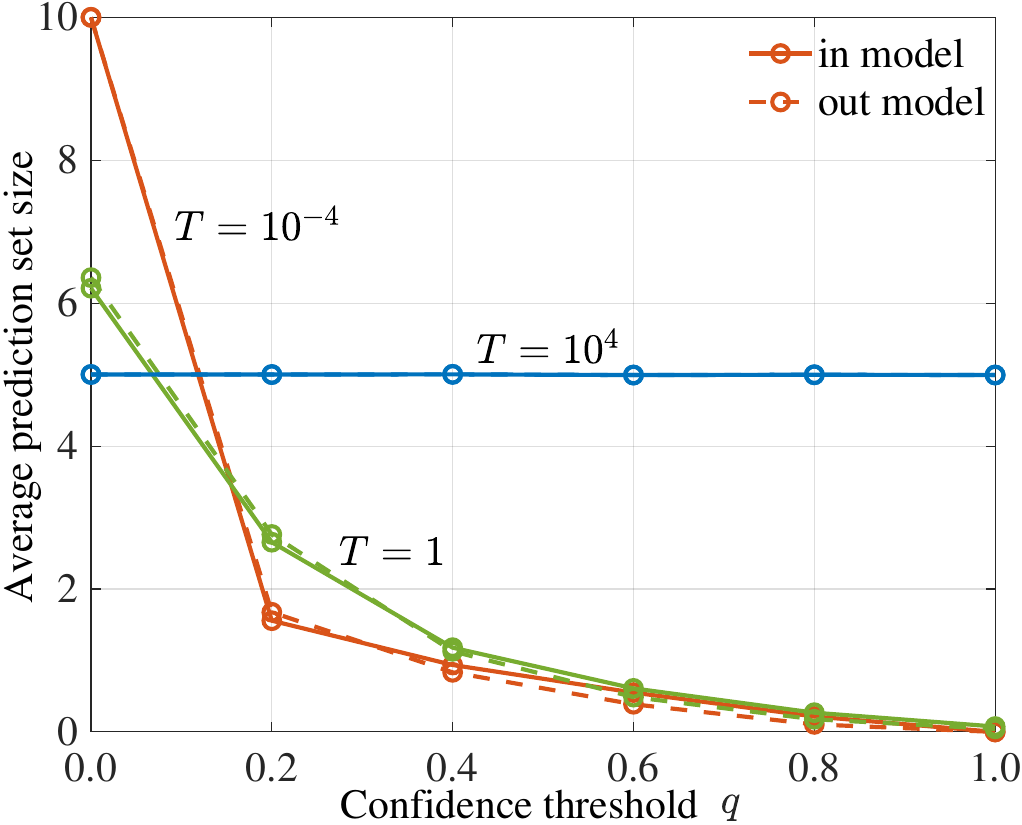}
    \includegraphics[width = 0.338\textwidth]{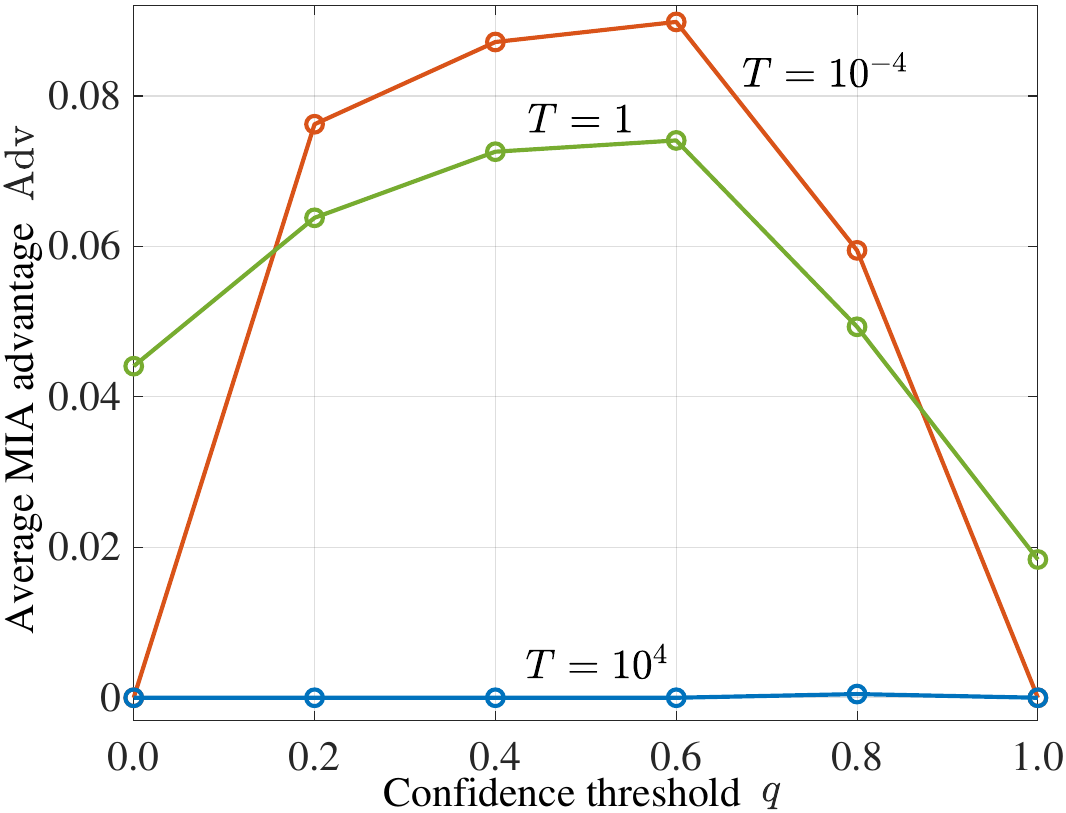}
    }
    \caption{Average prediction set size (left) and attacker's advantage (right) based on DS disclosure as a function of the confidence threshold $q$ with relative calibration error $\Delta=0.2$, aleatoric uncertainty $\epsilon_a=0.5$, epistemic uncertainty $\epsilon_e=0.25$.}\label{thre}
    \vspace{-5mm}
\end{figure*}
Finally, Fig. \ref{Fig_Adv_EU} shows the average attacker's advantage as a function of the epistemic uncertainty $\epsilon_e$ for $\Delta = 0.2$ and $\epsilon_a=0.5$. An increased epistemic uncertainty reflects smaller training sets, causing a larger uncertainty on the target model. Accordingly, the effectiveness of MIA decreases with $\epsilon_e$, becoming progressively lower due to the growing difficulty of the attacker to predict the model's outputs when including or not the target sample $(x_0,y_0)$ in the training data set.

\vspace{-0.3cm}
\subsection{Evaluating MIA with DS Disclosure}\label{subsec_DS}
In this final subsection, we further analyze the performance of the DS-observed attacker by studying the impact of the temperature parameter $T$ and of the confidence threshold $q$ in \eqref{obs_p_q_random}. As discussed in Sec. \ref{analysis_DS} (see also Fig. \ref{TV_thre_fig}), the bound \eqref{approx_ub_DS} provides a U-shaped curve for the attacker's advantage, as excessively small or large thresholds yield predicted sets that are nearly independent of the input $x_0$. In fact, as seen in Fig. \ref{thre} (left), as the threshold $q$ approaches $0$, the target model tends to produce a prediction set containing all labels, while with $q$ near $1$, it generates an empty prediction set. Neither extreme provides useful information for the attacker. The most effective attack occurs at a moderate $q$ value, here $q = 0.6$, where the prediction set is expected to yield the most informative insights for the attacker. This trend is confirmed by Fig. \ref{thre} (right), which shows the average attacker's advantage as a function of the threshold $q$ for different values of the temperature parameter $T$ in the randomized thresholding \eqref{p_thre}.

Finally, we observe that the attacker's advantage follows the analysis in Sec. \ref{analysis_DS}, illustrated in Fig. \ref{TV_thre_fig}, also in terms of its dependence on the randomness of the thresholding scheme \eqref{p_thre}. As the temperature parameter $T$ approaches $0$, the thresholding scheme becomes deterministic, providing the most information to the attacker for a given threshold $q$, and resulting in the highest advantage. Conversely, as $T$ grows, the thresholding noise is so large that the attacker can obtain vanishing information about the target model, and the advantage consistently attains the theoretical lower limit of $0$.

Further experimental results are provided in Appendix \ref{apdx_other_factor}, and discussions on performance under large-class settings are included in Appendix \ref{apdx_larg_class}.

\vspace{-3mm}
\section{Conclusions}\label{LiRA_conclusion}
This paper has introduced a theoretical framework to analyze the attacker's advantage in MIAs as a function of the model's calibration, as well as of the aleatoric and epistemic uncertainties. Following the theoretical framework of LiRA-style attacks, we regarded MIA as a hypothesis-testing problem; and we considered three typical types of observations that can be acquired by the attacker when querying the target model: confidence vector (CV), true label confidence (TLC), and decision set (DS). By making the flexible modeling assumption that the output confidence vectors follow the Dirichlet distributions, thus accounting for both aleatoric and epistemic uncertainties, we derived upper bounds on the attacker's advantage. The bounds yield analytical insights on the impact of the relative calibration error, aleatoric uncertainty, and epistemic uncertainty on the success of MIAs. For DS disclosure, we studied for the first time the impact of the threshold and of randomization on the effectiveness of MIA.

Our theoretical analysis is formulated based on the LiRA-style attacks that assume the worst-case scenario defined by the membership inference game in Definition \ref{def_MIG}. Accordingly, the attacker has accurate knowledge of the data distribution, as well as of the target model's architecture and training algorithm. We also do not impose constraints on computational resources, enabling the attacker to train enough shadow models to accurately estimate the output distributions for models trained with and without the target data point. This setting represents the most favorable conditions for the attacker, providing an upper bound on the MIA advantage.

Given that the derived bounds assume a worst-case adversary, the analysis in this paper is also applicable to recent LiRA-style attacks that inherently limit attack performance. These include settings with data shifts unknown to the attacker, as well as computational constraints \cite{ali2023membership, ye2022enhanced, zarifzadeh2023low}. Notably, reference \cite{zarifzadeh2023low} uses a Bayesian method to provide a lightweight version of LiRA-style attack, requiring only one or two shadow models.

For practical guidance on the design of protection strategies against LiRA-based MIAs derived from our theoretical analysis, please see Appendix \ref{apdx_prac_use}.

Future work may specialize the analysis to data domains such as text, images, and tabular data, as well as to model architectures such as transformers. Additionally, from the defender's perspective, it would be interesting to investigate the effectiveness of countermeasures to MIA such as regularization, model pruning, and data augmentation.

\begin{figure*}[t]
    \centering
    \setlength{\abovecaptionskip}{-2pt}
    {
    \includegraphics[width = 0.32\textwidth]{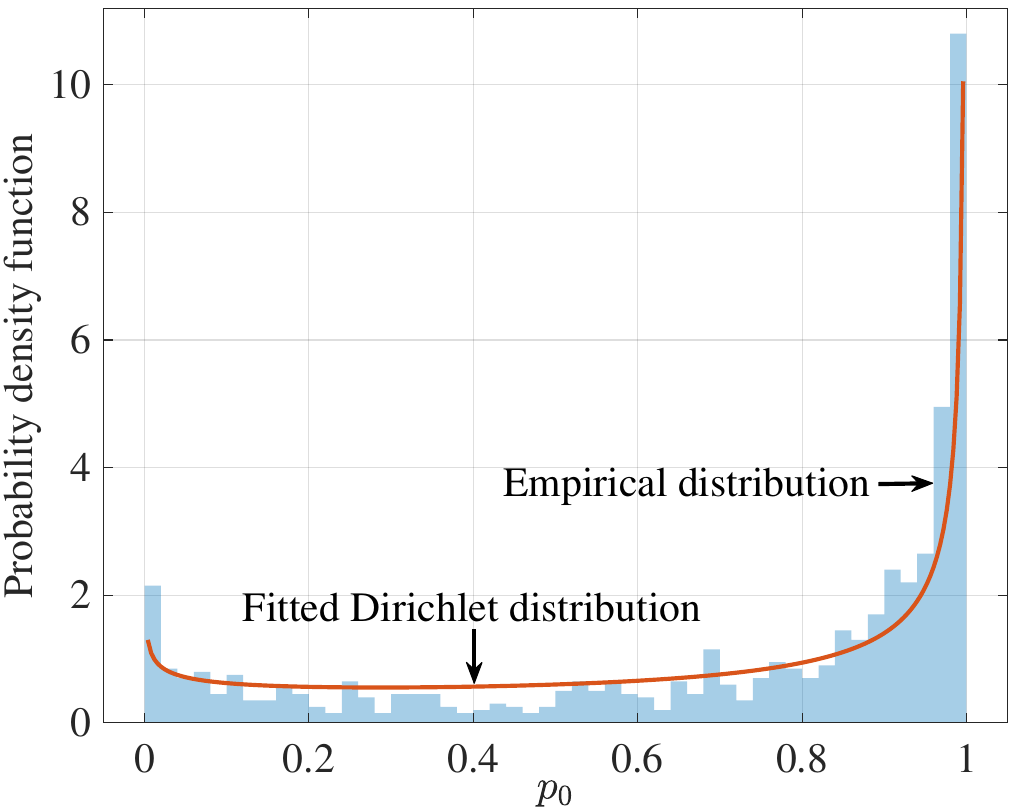}
    \includegraphics[width = 0.32\textwidth]{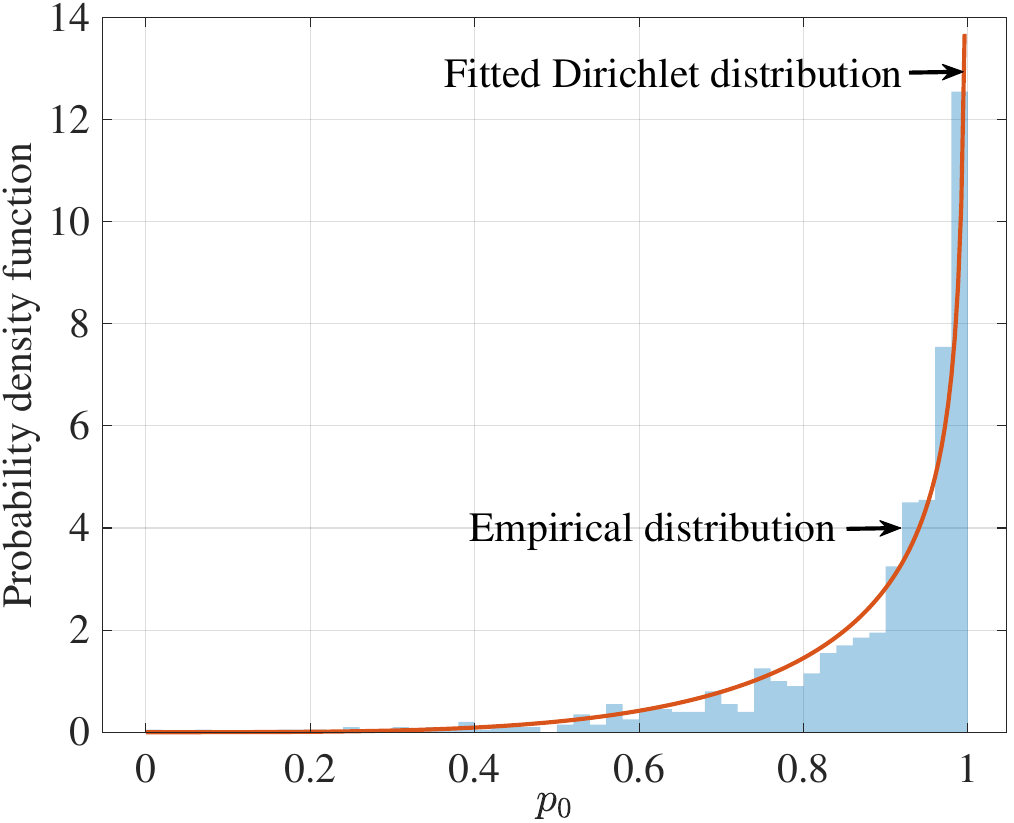}
    }
    \caption{Empirical distributions of the confidence level $p_0$ and corresponding marginal of the fitted Dirichlet distributions for models trained without (left) and with (right) the target sample $(x_0,y_0)$ for STL-10 data set with VGG-16.}\label{fitting2}
    \vspace{-5mm}
\end{figure*}

\vspace{-3mm}
\appendix
\subsection{Derivation of the Dirichlet Parameters in \eqref{gamma_out} and \eqref{gamma_in}}\label{Appendix_eq}

Equations \eqref{gamma_out} and \eqref{gamma_in}, which express the Dirichlet parameters for the distributions of $f^{\mathrm{out}}(p)$ and $f^{\mathrm{in}}(p)$ as functions of the aleatoric uncertainty $\epsilon_a$ in \eqref{AU}, epistemic uncertainty $\epsilon_e$ in \eqref{EU}, and relative calibration error $\Delta$ in \eqref{RCE}, are obtained as follows.

By substituting the aleatoric uncertainty $\epsilon_a$ from \eqref{AU} into the relative calibration error $\Delta$ in \eqref{RCE}, the expected confidence $\mathbb{E}_{f^{\mathrm{in}}}\left[p_0\right]$ can be expressed in terms of parameters $\epsilon_a$ and $\Delta$. Then, substituting this expected confidence and the epistemic uncertainty $\epsilon_e$ from \eqref{EU} into the average confidence level \eqref{average} with $z = \mathrm{in}$ and $k=0$, the parameter $\gamma^{\mathrm{in}}_0$ can be represented as a function of parameters $\Delta$, $\epsilon_a$, and $\epsilon_e$. Using the assumption $\gamma^{\mathrm{in}}_1=\ldots=\gamma^{\mathrm{in}}_{K-1}$ for all other candidate classes, we get $\gamma^{\mathrm{in}}_k$ for $k=1,2\ldots,K-1$ using \eqref{EU}. For the distribution $f^{\mathrm{out}}(p)$, the parameters $\gamma^{\mathrm{out}}_k$ can be obtained by setting $\Delta=0$ in the expression for $\gamma^{\mathrm{in}}$ in \eqref{gamma_in}.

\vspace{-3mm}
\subsection{Proof of Proposition \ref{Prop_1}}\label{Appendix_prop1}
The KL divergence between two Dirichlet distributions in \eqref{lower_bound2} can be expressed as \cite{lin2016dirichlet}
\begin{align}\label{Outer_bound_Dir}
    &D\big(f^{\mathrm{out}}(p;\gamma^{\mathrm{out}})\|f^{\mathrm{in}}(p;\gamma^{\mathrm{in}})\big)=\ln\frac{B(\gamma^{\mathrm{in}})}{B(\gamma^{\mathrm{out}})}\nonumber\\
    &\qquad+\sum_{k=0}^{K-1}\big(\gamma^{\mathrm{out}}_k-\gamma^{\mathrm{in}}_k\big)\Big(\psi(\gamma^{\mathrm{out}}_k)-\psi\Big(\sum_{i=0}^{K-1} \gamma^{\mathrm{out}}_i\Big)\Big),
\end{align}
where we have defined that digamma function $\psi(x)=\Gamma'(x)/\Gamma(x)$, multivariate Beta function $B(\gamma^\mathrm{out/in})=\prod_{i=0}^{K-1}\Gamma(\gamma^\mathrm{out/in}_i)\big/\Gamma(\sum_{i=0}^{K-1} \gamma^\mathrm{out/in}_i)$ and Gamma function $\Gamma(x)=\int_0^{+\infty}t^{x-1}e^{-t}dt$. The same logic applies for $D \big( f^{\mathrm{in}} (p; \gamma^{\mathrm{in}}) \| f^{\mathrm{out}} (p; \gamma^{\mathrm{out}})\big)$.

Accordingly, the square root of the symmetrized KL divergence in the approximate bound \eqref{lower_bound2} in Lemma \ref{lemma_ub} given CV observations can be calculated as
\begin{align}\label{approx_KL_append}
    \mathrm{Adv}^{\mathrm{CV}}_{\alpha} \leq& \Big(D\big(f^{\mathrm{out}}\left(p;\gamma^{\mathrm{out}}\right) \| f^{\mathrm{in}}\left(p;\gamma^{\mathrm{in}}\right)\big) \nonumber\\
    &+ D\big(f^{\mathrm{in}}\left(p; \gamma^{\mathrm{in}}\right) \| f^{\mathrm{out}}\left(p; \gamma^{\mathrm{out}}\right)\big)\Big)^{\frac{1}{2}}\nonumber\\
    \triangleq &\mathrm{Adv}_{\alpha}^{\mathrm{CV-ub}},
\end{align}
by using the definitions of $\gamma_k^{\mathrm{out}}$ in \eqref{gamma_out} and of $\gamma_k^{\mathrm{in}}$ in \eqref{gamma_in}.

\vspace{-3mm}
\subsection{Proof of Proposition \ref{Prop_3}}\label{Appendix_prop3}
We leverage the strong data processing inequality \cite[Proposition \uppercase\expandafter{\romannumeral2}.4.10]{cohen1998comparisons}\cite{he2024information} to obtain the upper bound of the MIA advantage given DS observations as
\begin{align}\label{KL_TV}
    \frac{D\big(f^{\mathrm{out}}(b)\|f^{\mathrm{in}}(b)\big)}{D\big(f^{\mathrm{out}}(p)\|f^{\mathrm{in}}(p)\big)}&\leq \max_{p,p'} \left\| f(b|p)-f(b|p')\right\|_{\mathrm{TV}} \nonumber\\
    &\triangleq \delta^2_{T,q},
\end{align}
where $\|P-Q\|_{\mathrm{TV}}$ denotes the total variance (TV) distance between distributions $P$ and $Q$. Note that we have the inequality $\delta_{T,q}^2\leq 1$. Accordingly, the advantage of the attacks given DS observation can be upper bounded as
\begin{align}\label{KL_TV2}
    \mathrm{Adv}^{\mathrm{DS}}_{\alpha}&\leq \sqrt{D\big(f^{\mathrm{out}}(b)\|f^{\mathrm{in}}(b)\big) + D\big(f^{\mathrm{in}}(b)\|f^{\mathrm{out}}(b)\big)}\nonumber\\
    &=\delta_{T,q}\cdot\mathrm{Adv}^{\mathrm{CV-ub}}_{\alpha},
\end{align}
where we have used Lemma \ref{lemma_ub} and \eqref{KL_TV}.

\vspace{-3mm}
\subsection{Additional Validation of the Dirichlet Assumption}\label{Appendix_fitting}
To further validate the use of a Dirichlet distribution to model the confidence probability, we conduct additional experiments on the STL-10 data set using the pre-trained VGG-16 model. We randomly select $N^{\mathrm{tr}}=4000$ images for data sets $\mathcal{D}^\mathrm{out}$ and $\mathcal{D}^{\mathrm{in}}$. The VGG-16 model is fine-tuned using the SGD optimizer with a learning rate of $0.001$, momentum of $0.9$, and weight decay of $0.0005$, over $50$ epochs.
As per the methodology in Sec. \ref{subsec_Dir}, we obtain probability vectors for images with the ground truth label $0$ from both the training and test sets, fitting these to Dirichlet distributions using maximum likelihood estimation with the L-BFGS-B algorithm.

Fig. \ref{fitting2} shows the fitted and empirical distribution probability density functions for the target sample $(x_0,y_0)$, confirming the flexibility of the Dirichlet assumption for modeling confidence vectors.

\begin{figure*}[t]
    \centering
    \setlength{\abovecaptionskip}{-2pt}
    {\includegraphics[width = 0.32\textwidth]{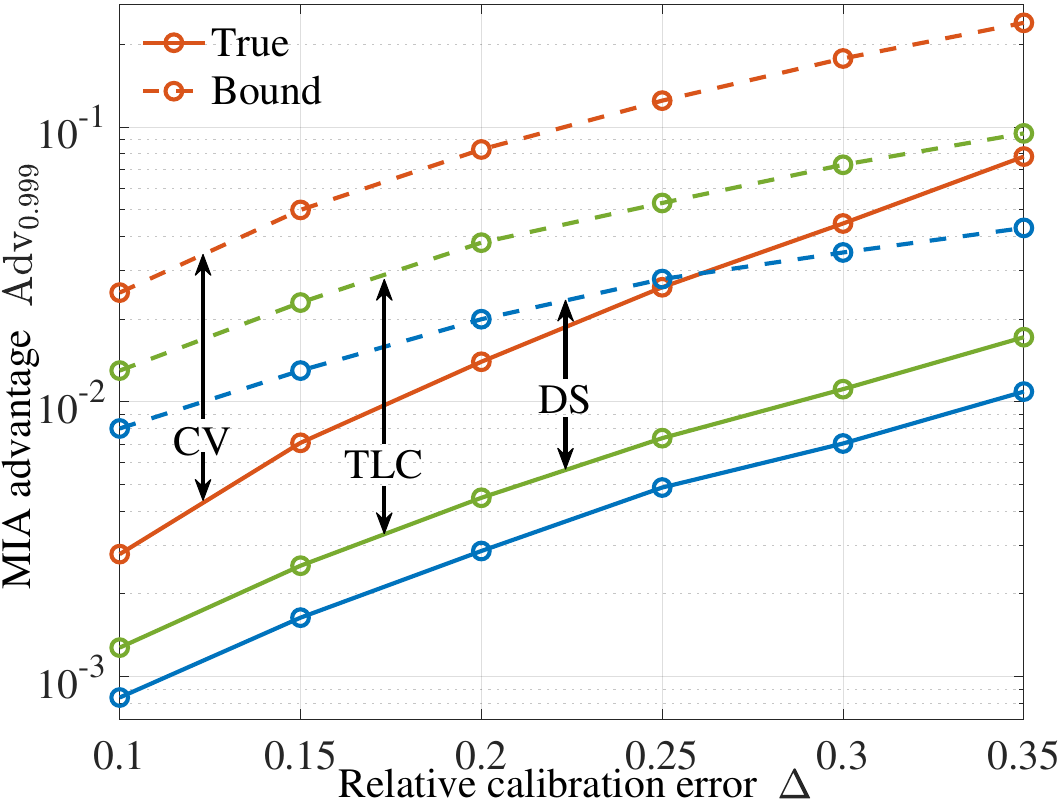}
    \includegraphics[width = 0.318\textwidth]{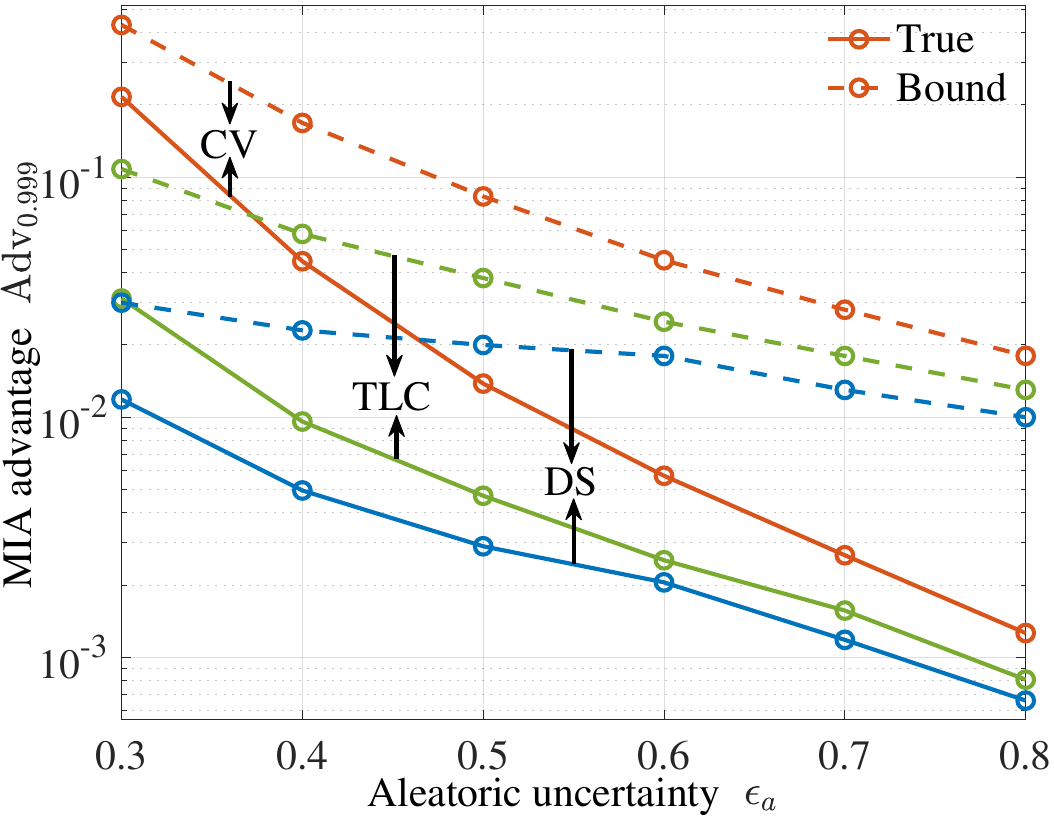}
    \includegraphics[width = 0.32\textwidth]{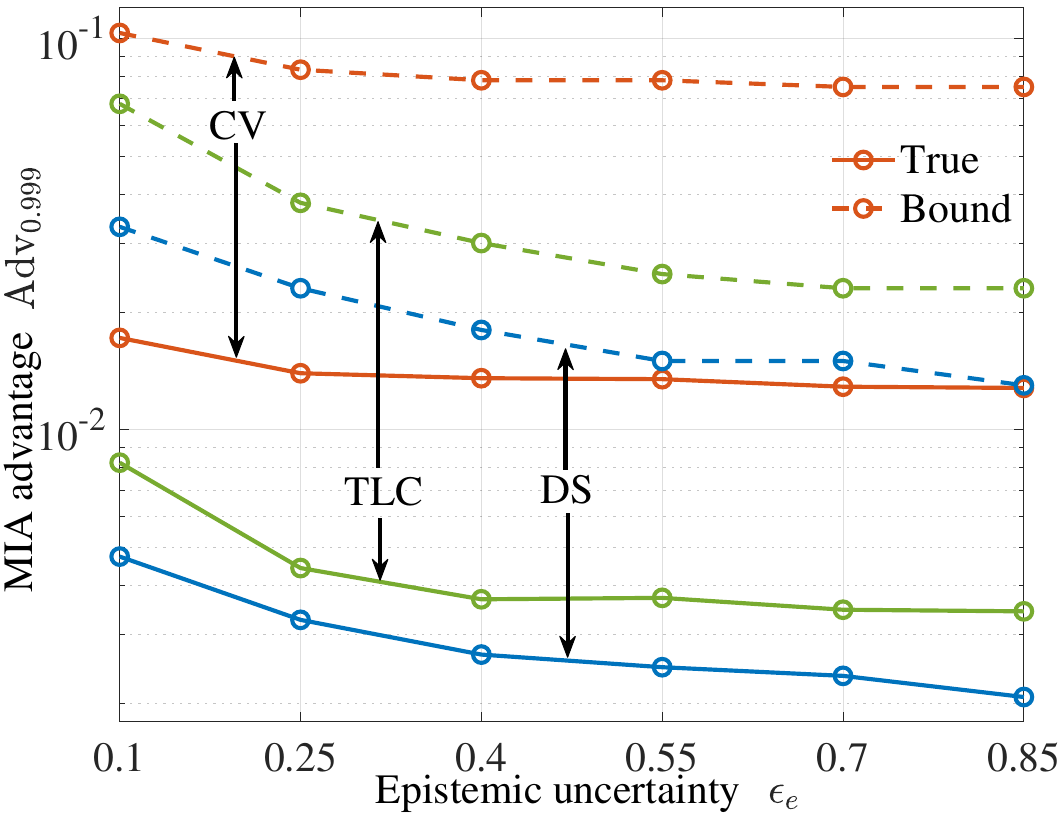}
    }
    \caption{True advantage \eqref{eq_Adv} and bounds \eqref{lower_bound2} for the attacker's advantage with CV, TLC, and DS observations in high TNR regime, $\alpha = 0.999$, as a function of calibration error, $\Delta$ (left); of aleatoric uncertainty, $\epsilon_a$ (middle); and of epistemic uncertainty, $\epsilon_e$ (right). }\label{fig_RCE_AU_EU_highTNR}
    \vspace{-5mm}
\end{figure*}

\begin{figure*}[t]
    \centering
    \setlength{\abovecaptionskip}{-2pt}
    {\includegraphics[width = 0.312\textwidth]{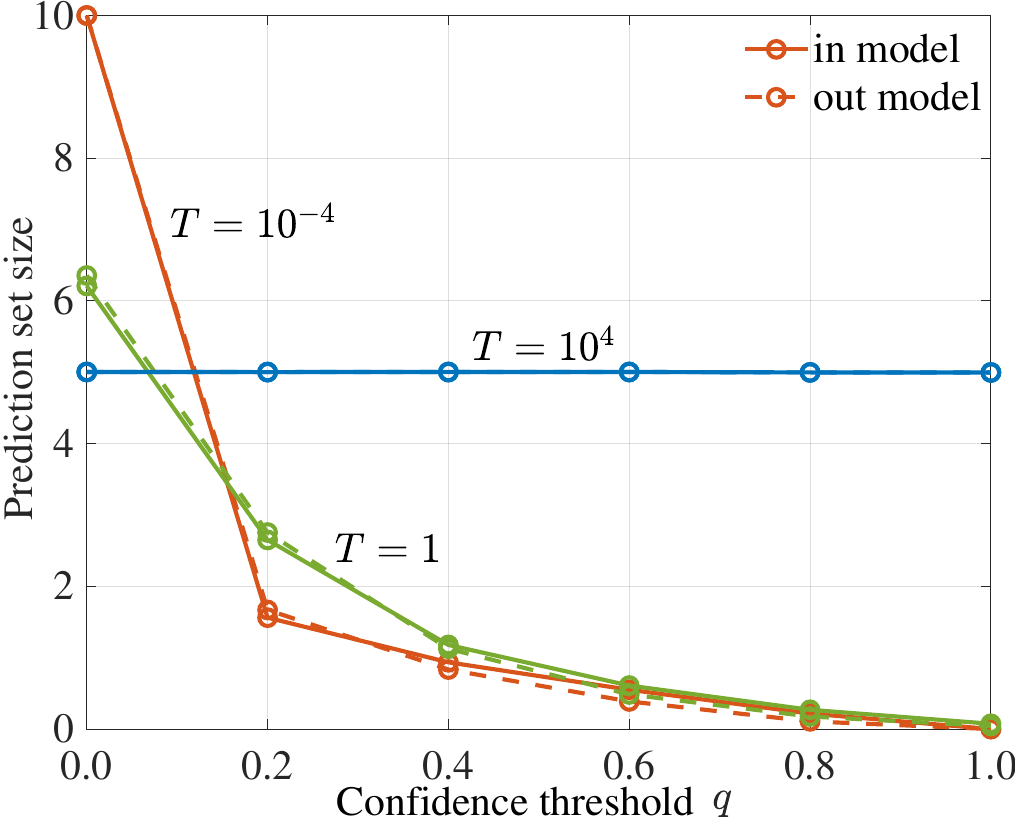}
    \includegraphics[width = 0.32\textwidth]{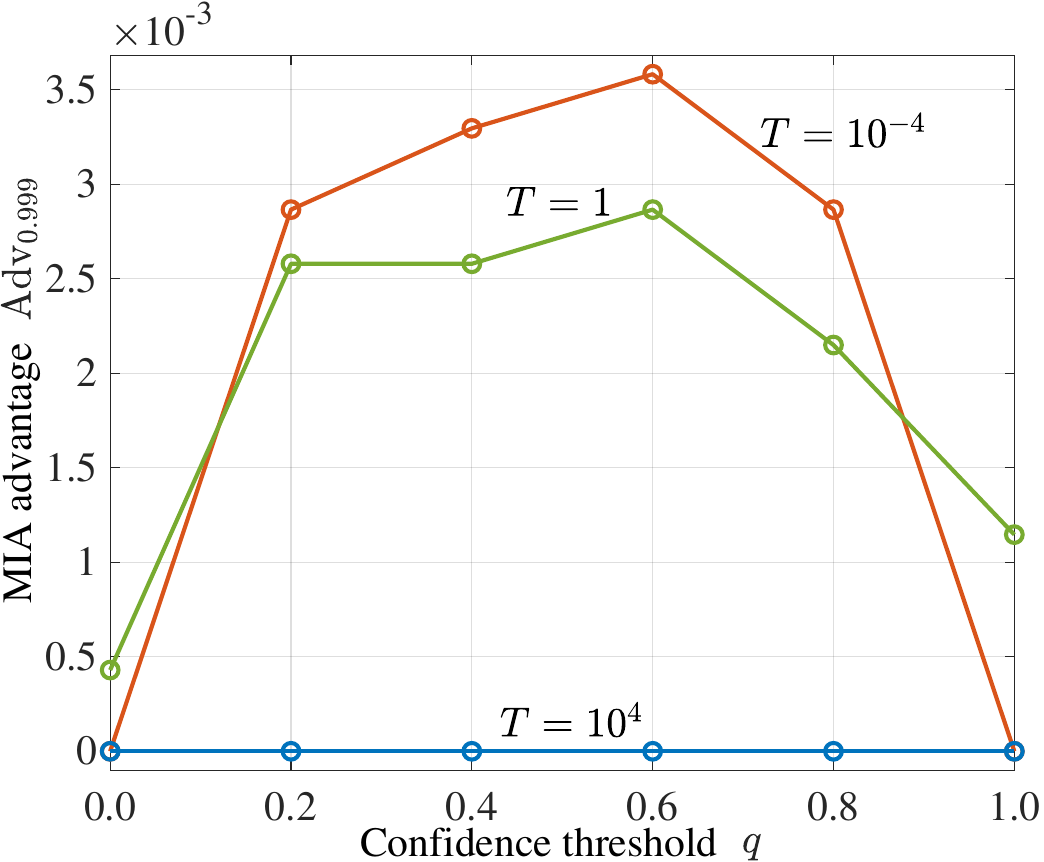}
    }
    \caption{Prediction set size (left) and attacker's advantage (right) based on DS disclosure in high TNR regime, $\alpha = 0.999$, as a function of the confidence threshold $q$.} \label{fig_thre_highTNR}
    \vspace{-5mm}
\end{figure*}

\vspace{-3mm}
\subsection{Performance in High TNR Regime}\label{Appendix_highTNR}
In Sections \ref{subsec_CE_AU_EU} and \ref{subsec_DS}, we have evaluated the average MIA advantage, $\mathrm{Adv}$, across all TNR values for a broad assessment. As a complementary analysis, we further focus on high TNR to gain deeper insights into the attack's performance in privacy-sensitive applications. Evaluating the advantage at high TNR is particularly meaningful in these scenarios because it reflects the attacker's ability to correctly identify non-training data while still effectively detecting training data. This ensures that the attack is both precise and reliable.

Accordingly, we evaluate the attack performance $\mathrm{Adv}_{0.999} = 0.999-\beta_{0.999}$ at high TNR ($\alpha = 0.999$), using the same settings as in Sec. \ref{subsec_CE_AU_EU} and Sec. \ref{subsec_DS}. Fig. \ref{fig_RCE_AU_EU_highTNR} shows the attacker's advantage with CV, TLC, and DS observations when $\alpha = 0.999$ as a function of calibration error ($\Delta$), aleatoric uncertainty ($\epsilon_a$), and epistemic uncertainty ($\epsilon_e$). Fig. \ref{fig_thre_highTNR} illustrates the prediction set size and attacker's advantage for DS disclosure when $\alpha = 0.999$ as a function of the confidence threshold ($q$).

The results at high TNR are largely consistent with those averaged across all TNR values, highlighting the robustness of our theoretical analysis.

\vspace{-3mm}
\subsection{Defense Strategies Inspired by Our Analysis}\label{apdx_prac_use}
The theoretical analysis presented in this work offers insights into the impact of different sources of uncertainty and of different types of disclosures on MIA performance. These insights motivate practical defense strategies, which are outlined in the following.

\subsubsection{Calibration error, epistemic uncertainty, and aleatoric uncertainty}
Our analysis shows that smaller calibration error, higher aleatoric uncertainty, and higher epistemic uncertainty independently reduce MIA performance.
\begin{itemize}
    \item \textit{Calibration error:} Regularization-based defenses, such as L2 regularization \cite{shokri2017membership, choquette2021label}, dropout \cite{kaya2020effectiveness}, pruning \cite{wang2020against}, and model stacking \cite{salem2018ml}, reduce overfitting, which helps lower the model's confidence in its predictions. This makes the model more robust to MIAs by diminishing the risk of overconfidence, which attackers can exploit. Furthermore, to directly reduce calibration error, methods such as temperature scaling \cite{zade2025automatic}, Platt scaling, and isotonic regression \cite{guo2017calibration} can be used to reduce the model's overconfidence.
    \item \textit{Aleatoric uncertainty:} Defenses like adding perturbations to the original data \cite{hullermeier2021aleatoric, guo2007sensitivity, wang2019miasec}, using adversarial optimization to generate adversarial samples \cite{yang2023purifier}, and using generative models to create synthetic data \cite{hu2022defending} all increase aleatoric uncertainty. By introducing more noise into the data, these methods reduce the model's reliance on specific training data points, making it harder for attackers to infer membership.
    \item \textit{Epistemic uncertainty:} Techniques such as adversarial noise \cite{jia2019memguard}, differential privacy \cite{yeom2018privacy, choquette2021label, jayaraman2019evaluating, jayaraman2020revisiting}, deep ensembles and Bayesian learning \cite{hullermeier2021aleatoric, guo2007sensitivity} can increase epistemic uncertainty by introducing noise into the model's predictions or parameters. This reduces the model's predictability, hindering attackers' ability to make accurate inferences.
\end{itemize}

\subsubsection{CV, TLC, and DS disclosure-based schemes}
Our theoretical findings show that MIA advantage decreases as the prediction output becomes less informative, going from CV disclosure to TLC disclosure, and finally to DS disclosure. Accordingly, to enhance privacy, one may consider the following approaches:
\begin{itemize}
    \item \textit{Limit confidence disclosures:}  Avoid exposing the full CV. Instead, limiting disclosure to the TLC or DS provides better protection to MIAs. Models that generate a prediction set, such as set-valued predictions \cite{chzhen2021set}, random forests \cite{devetyarov2010prediction}, and conformal prediction models \cite{vovk2005algorithmic, angelopoulos2021gentle, cohen2023calibrating}, are thus inherently safer than traditional models that provide full probability distributions. Conformal prediction, in particular, is promising as it ensures both reliability and protection against MIAs.
    \item \textit{Enhance privacy in DS disclosure:} The benefits of DS disclosure in terms of protection against MIAs can be further enhanced through randomization techniques and threshold adjustments. For instance, temperature-based randomization in the softmax function introduces variability into the DS generation process, making it harder for attackers to determine membership with high certainty.
\end{itemize}

\begin{figure*}[t]
    \centering
    \setlength{\abovecaptionskip}{-2pt}
    {
    \includegraphics[width = 0.32\textwidth]{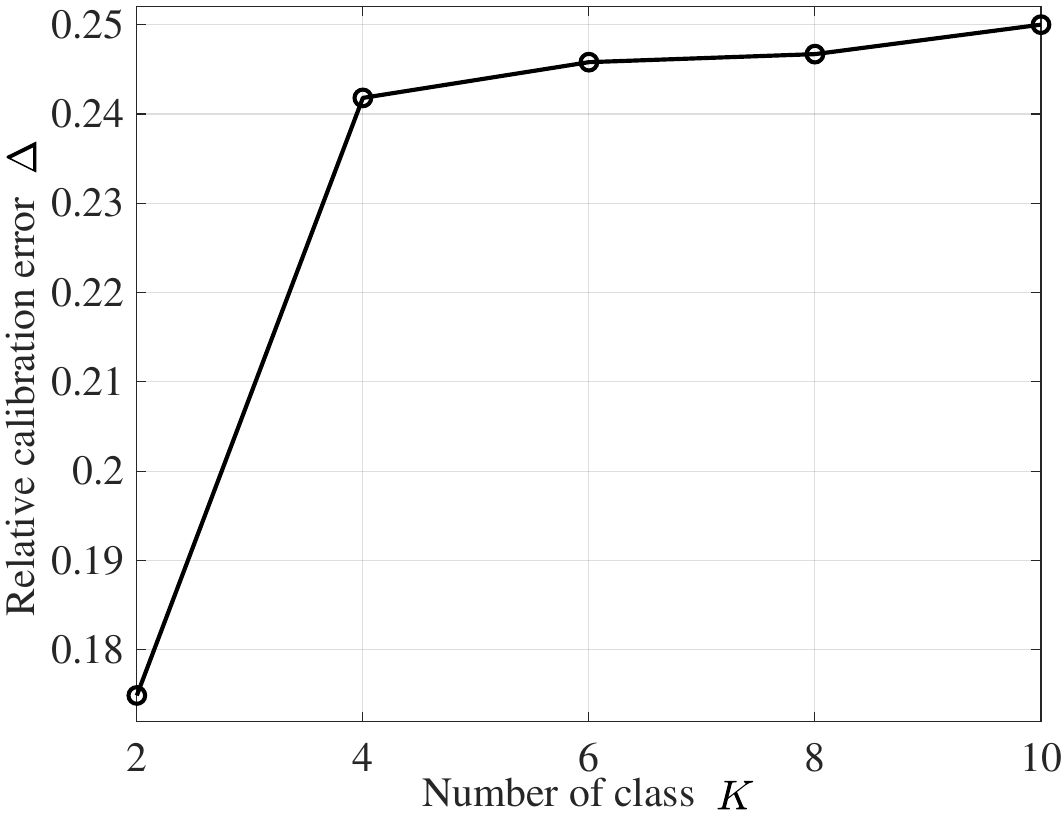}
    \includegraphics[width = 0.32\textwidth]{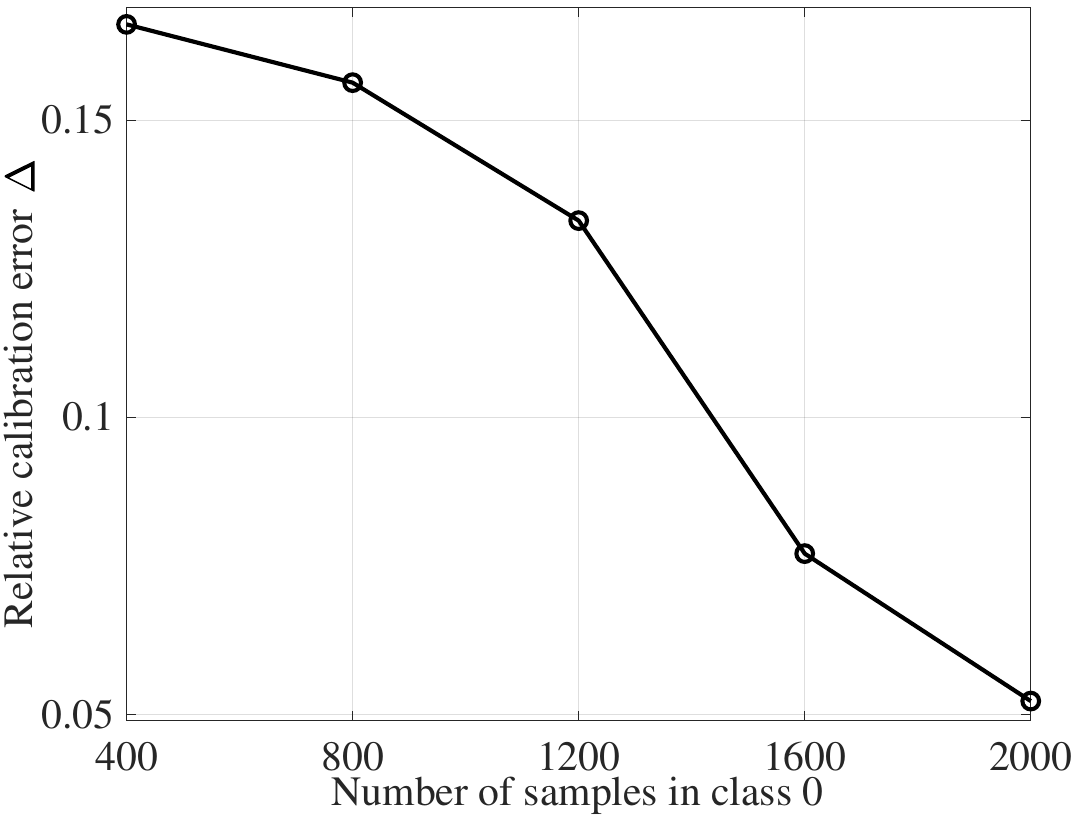}
    \includegraphics[width = 0.32\textwidth]{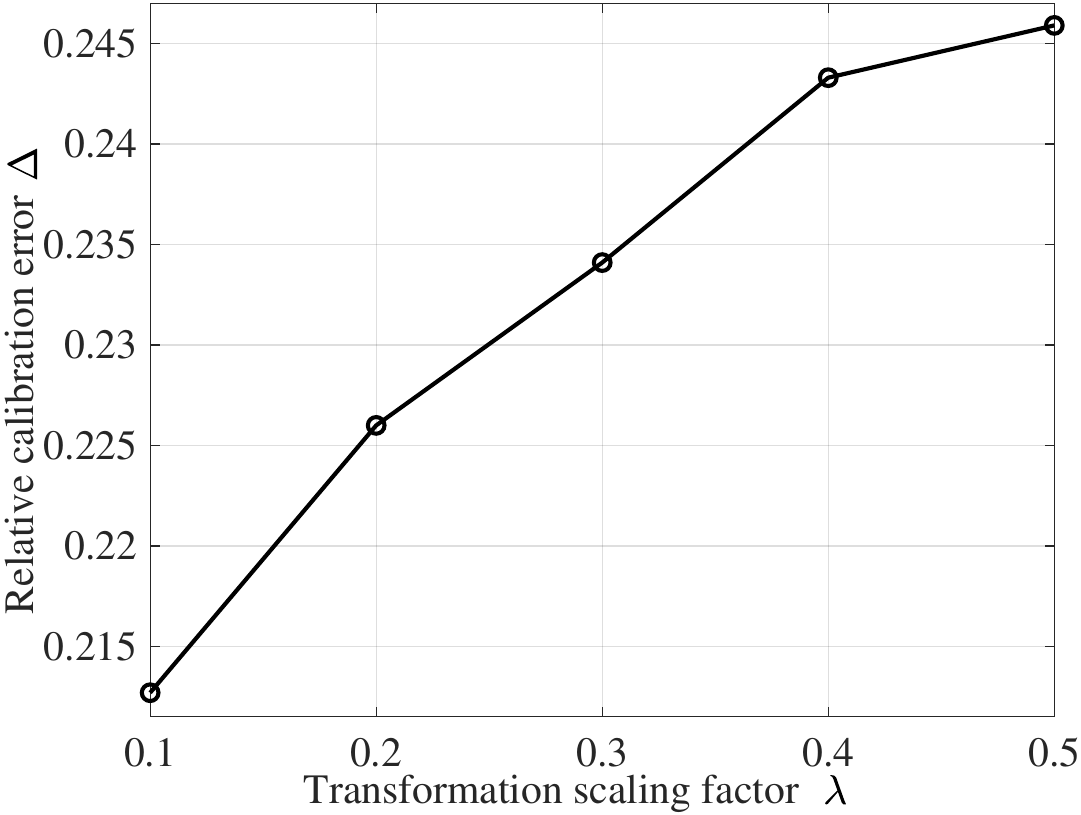}
    }
    \caption{The impact of the number of classes (left), of the subgroup size (middle), and of the subgroup distribution on model calibration (right).} \label{Fig_diff_num_cla}
    \vspace{-5mm}
\end{figure*}

\vspace{-3mm}
\subsection{Further Experimental Results}\label{apdx_other_factor}
This section demonstrates that factors previously linked to MIA vulnerability, such as class divisions \cite{shokri2017membership, salem2018ml, truex2019demystifying} and subgroup characteristics \cite{kulynych2019disparate}, are inherently captured by our theoretical framework through their impact on the  relative calibration error.

\subsubsection{Number of class divisions}
Existing empirical studies have shown that increasing the number of classes in a target model makes it more vulnerable to MIAs \cite{shokri2017membership, salem2018ml, truex2019demystifying}. As explained in \cite{shokri2017membership}, a larger number of output classes increases the dimensionality of the model's output vectors, exposing more internal information and making membership inference more effective. Additionally, with a fixed training data set size, a larger number of classes results in fewer samples per class, leading to greater overfitting and, consequently, increased susceptibility to MIAs \cite{shokri2017membership}. Another perspective, presented in \cite{truex2019demystifying}, attributes this effect to the partitioning of the input space into a larger number of smaller regions. This tighter segmentation of decision boundaries increases the influence of individual training samples on model predictions, thereby facilitating membership inference.

To evaluate the impact of class divisions within our theoretical framework, we conduct experiments on synthetic classification data sets with varying numbers of classes, $K = \left\{2, 4, 6, 8, 10\right\}$. Each data point consists of a $20\times 1$ feature vector sampled from a $K$-class Gaussian mixture model, under which data in each class follows a Gaussian distribution with mean vector draw from $\mathcal{N}(0, 25\mathbf{I})$, and a uniform standard deviation of $1$. To simulate label noise, we set the ground-truth probability $p^*_0 = 0.8$ for class $0$ by corrupting $20\%$ of its labels. This is done by randomly flipping label $0$ to other classes. The total number of training data points remains fixed at $200$, regardless of the number of classes.

We use a multilayer perceptron with a $20\times 1$ input, a $32$-unit hidden layer using ReLU activation, and a softmax output layer of size $K$. The model is trained for $50$ epochs using the Adam optimizer with a learning rate of $0.0005$, minimizing cross-entropy loss between predictions and noisy labels.

We evaluate the relative calibration error as defined in \eqref{RCE} of our paper. Specifically, the expected confidence $\mathbb{E}_{f^{\mathrm{in}}}\left[p_0\right]$ is calculated by averaging the model's predicted probabilities on training data with uncorrupted labels. As shown in Fig. \ref{Fig_diff_num_cla} (left), the relative calibration error $\Delta$ increases with the number of classes. Our theoretical analysis, consistent with Fig. \ref{Fig_Adv_RCE} in our paper, shows that a higher calibration error leads to a more pronounced MIA advantage. This explains why increasing class divisions improves MIA performance: it raises the model's calibration error, aligning with empirical findings in \cite{shokri2017membership}. With a fixed training data set size, more classes mean fewer samples per class, which, given sufficient model capacity and training resources, increases overfitting and calibration error. This, in turn, increases the vulnerability of the model to attacks, supporting our theoretical conclusions.

\subsubsection{Size and distribution of subgroups within the data}
Previous work \cite{kulynych2019disparate} has empirically analyzed the vulnerability of different subgroups in the training data to MIAs. The findings show that underrepresented groups, with fewer samples, are more vulnerable than well-represented groups, with more samples. Furthermore, when the feature distribution of a minority subgroup significantly deviates from that of the majority, the subgroup becomes more susceptible to attacks, facilitating membership inference.

To analyze the underlying causes of the above empirical observations within our theoretical framework,
we conduct experiments on the MNIST data set, treating class $0$ as a distinct subgroup and classes $1$ to $9$ as the majority group. The classification model is a CNN with two convolutional layers ($32$ and $64$ filters) followed by two fully connected layers ($128$ and $10$ units). We train the model using the Adam optimizer with a learning rate of $0.001$ for $20$ epochs, minimizing cross-entropy loss. To simulate the aleatoric uncertainty, we set the ground-truth probability for class $0$ to $p^*_0=0.8$ and introduce label noise as in the previous example. The relative calibration error is then evaluated using equation \eqref{RCE} from our paper.

To analyze the effect of subgroup size on MIA performance, we vary the size of samples in class $0$ in the set $\{400, 800, 1200, 1600, 2000\}$, while keeping $2000$ samples per class for all other classes. As shown in Fig. \ref{Fig_diff_num_cla} (middle), the relative calibration error $\Delta$ decreases as the size of the subgroup for class $0$ increases. This is because the model overfits more easily on smaller subgroups, learning only limited patterns from the data, which increases calibration error. As the sample size of class $0$ increases, the calibration error decreases.

To assess the impact of subgroup distribution on MIA performance, we design an experiment with fixed class sizes of $400$ samples per class on the MNIST data set. For class $0$, we induce controlled distribution shifts using affine transformations, including rotation, translation, scaling, and shearing. The transformation intensity is adjusted by a scaling factor $\lambda \in \left\{0.1, 0.2, 0.3, 0.4, 0.5\right\}$, which linearly determines the magnitude of each transformation: rotation up to $\pm 40\lambda^\circ$, translation by $±20\lambda\%$ of the image size, scaling from $1.0$ to $1.0+0.5\lambda$, and shearing up to $\pm20\lambda^\circ$. This controlled variation progressively increases the distributional divergence between the modified class $0$ and unmodified classes $1-9$, allowing a quantitative analysis of how subgroup distribution influences MIA performance. As shown in Fig. \ref{Fig_diff_num_cla} (right), the relative calibration error increases as the feature distribution of class 0 deviates further from the majority classes due to stronger affine transformations. The more distinct the subgroup's distribution, the more the model memorizes its unique patterns, leading to overfitting and higher calibration error. This effect is more pronounced in smaller subgroups, where limited data prevents the model from generalizing effectively.

In summary, our analysis shows that the number of class divisions, as well as the size and distribution of subgroups, primarily influence the model's relative calibration error. Instead of modeling these factors separately, our study explicitly formulates relative calibration error to examine its impact on MIA advantage, providing a more fundamental understanding of the issue. Since these factors inherently contribute to calibration error, our unified modeling approach effectively captures their effects within a single framework.

\vspace{-3mm}
\subsection{Performance under Large-Class Settings} \label{apdx_larg_class}

\begin{figure*}[t]
    \centering
    \setlength{\abovecaptionskip}{-2pt}
    {
    \includegraphics[width = 0.32\textwidth]{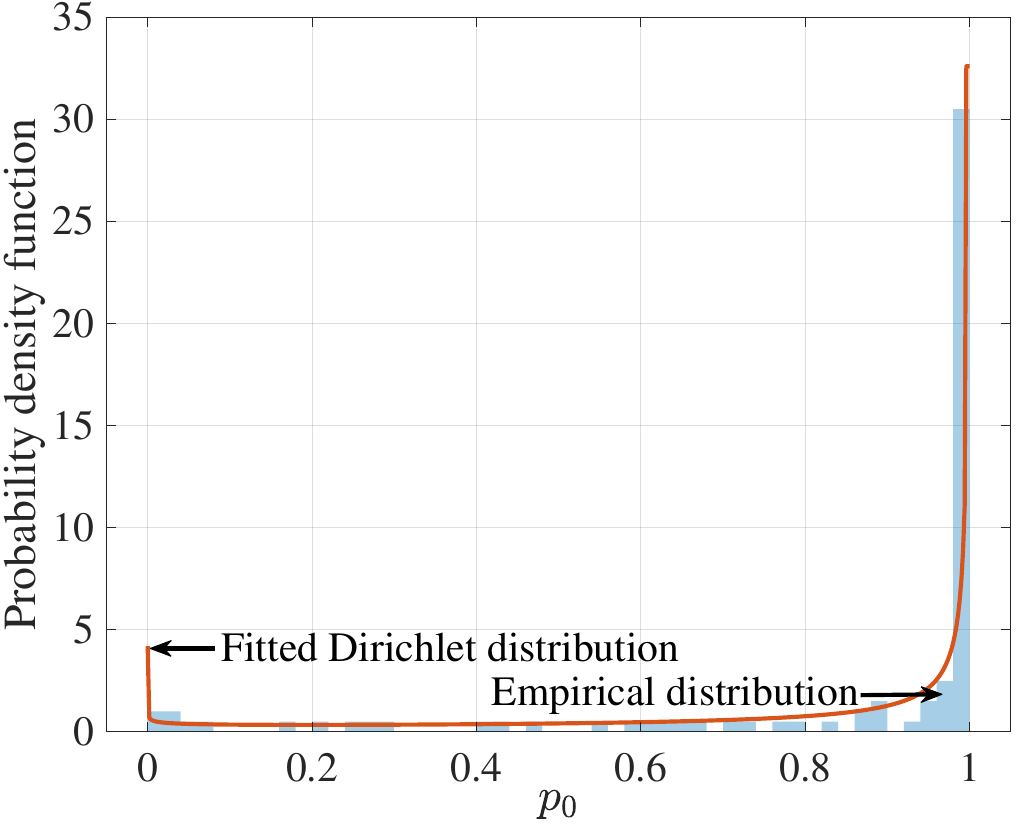}
    \includegraphics[width = 0.32\textwidth]{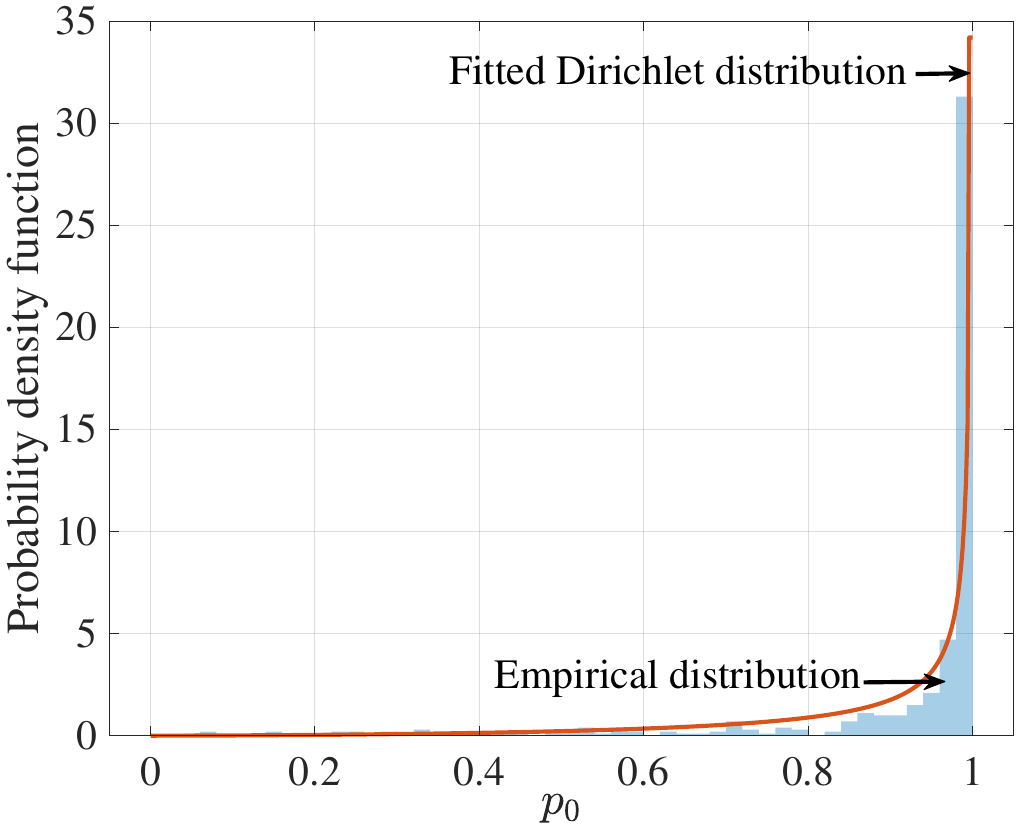}
    }
    \caption{Empirical distributions of the confidence level $p_0$ and corresponding marginal of the fitted Dirichlet distributions for models trained without (left) and with (right) the target sample $(x_0,y_0)$ for CIFAR-100 data set with ResNet-50.}\label{fitting3}
    \vspace{-3mm}
\end{figure*}

\begin{figure*}[t]
    \centering
    \setlength{\abovecaptionskip}{-2pt}
    {\includegraphics[width = 0.32\textwidth]{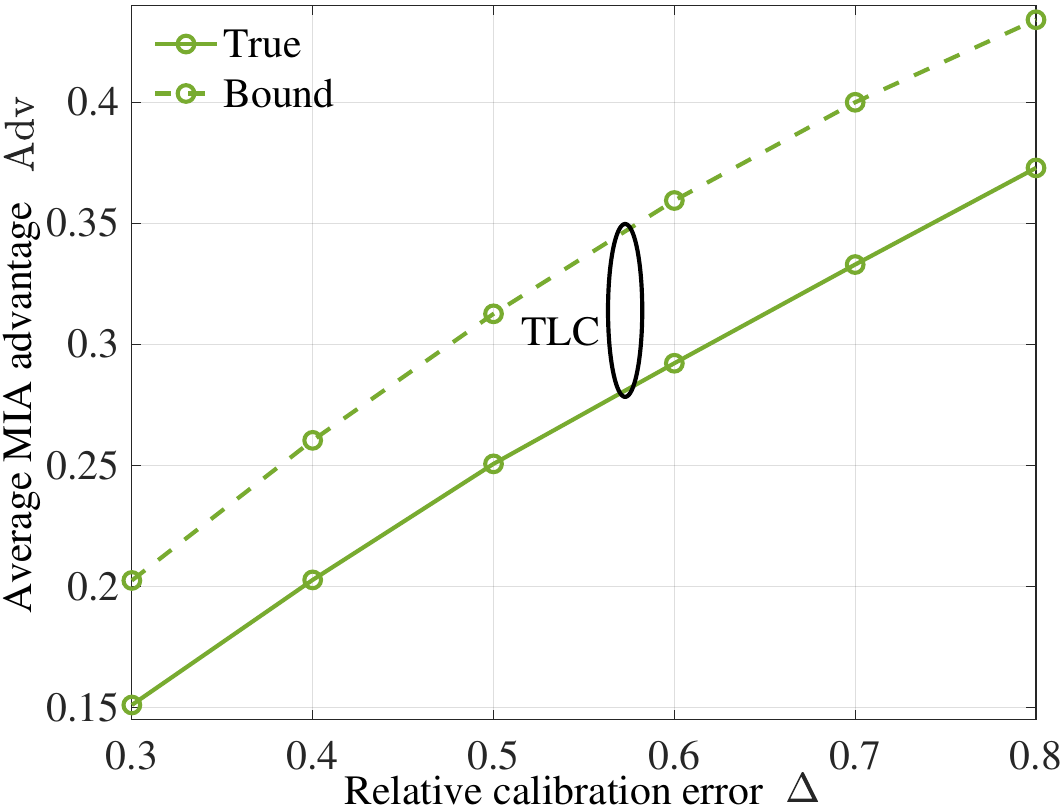}
    \includegraphics[width = 0.32\textwidth]{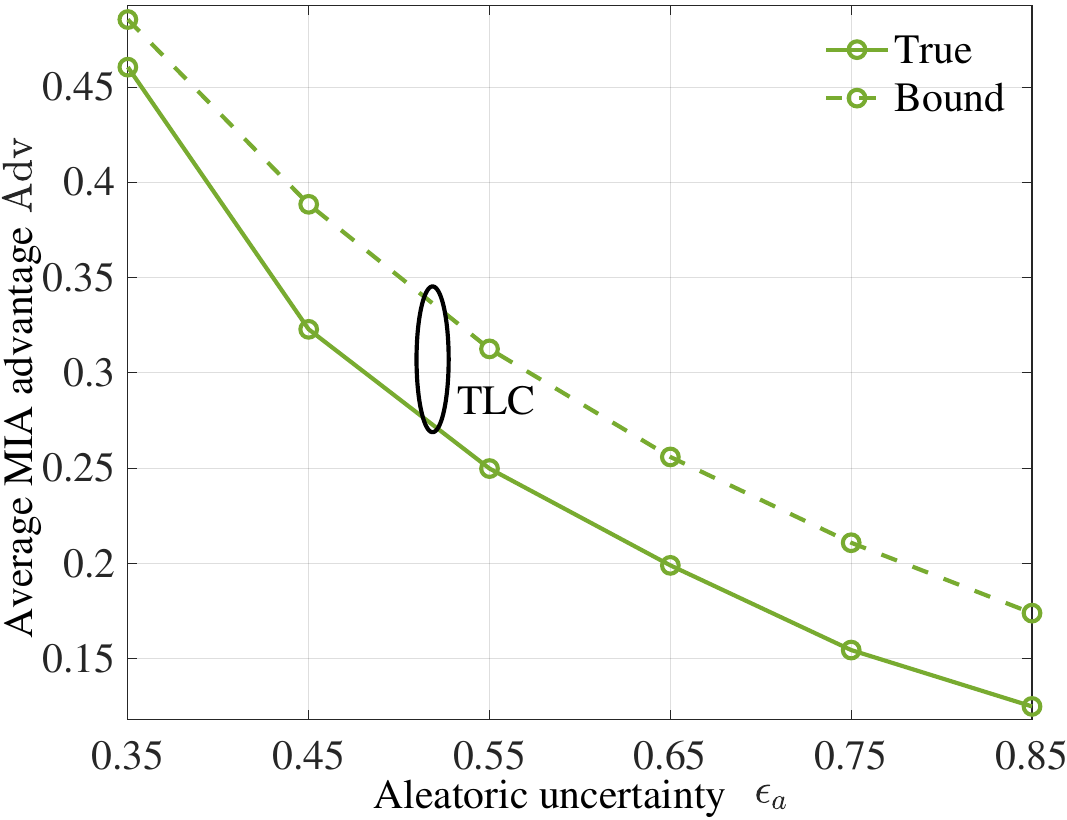}
    \includegraphics[width = 0.32\textwidth]{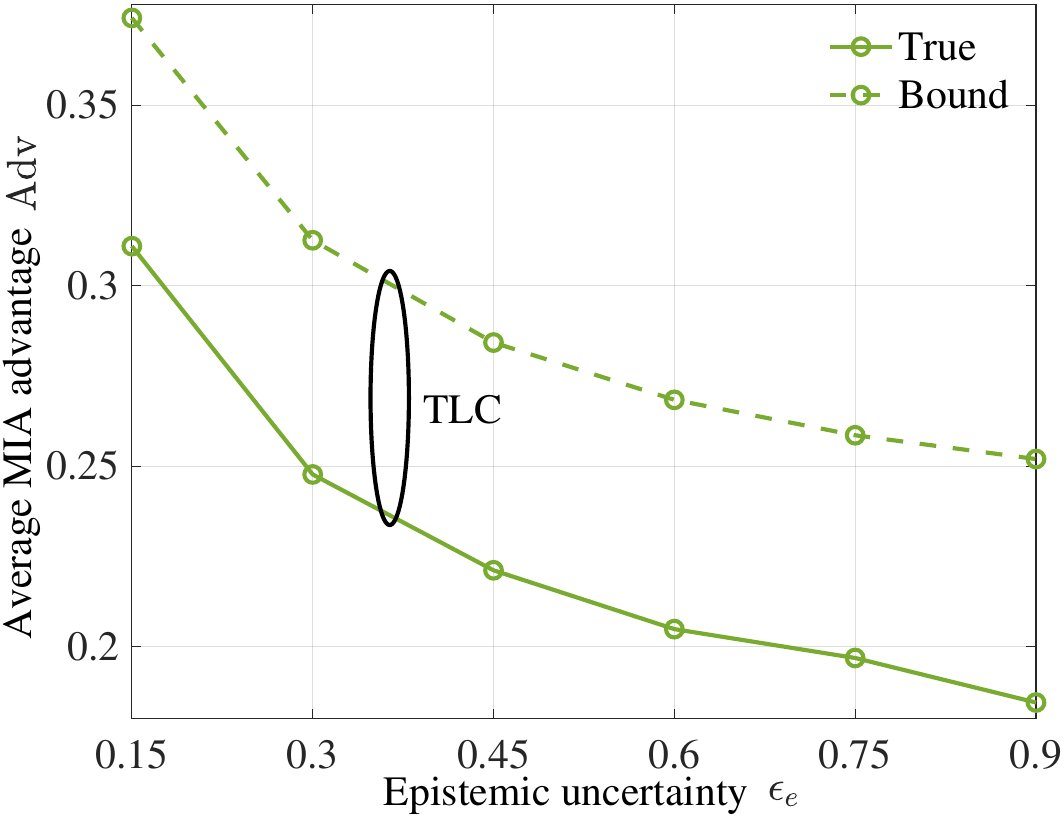}
    }
    \caption{True advantage \eqref{eq_Adv} and bounds \eqref{lower_bound2} for the attacker's advantage with TLC disclosure as a function of calibration error, $\Delta$ (left, $\epsilon_a = 0.55$, $\epsilon_e = 0.3$, $q = 0.2$); of aleatoric uncertainty, $\epsilon_a$ (middle, $\Delta = 0.5$, $\epsilon_e = 0.3$, $q = 0.2$); and of epistemic uncertainty, $\epsilon_e$ (right, $\Delta = 0.5$, $\epsilon_a = 0.55$, $q = 0.4$).} \label{fig_RCE_AU_EU_large_scale}
    \vspace{-5mm}
\end{figure*}

Note that empirical evaluation under CV and DS disclosures becomes infeasible when the number of classes is large, due to the challenge of high-dimensional distribution estimation. This is an inherent limitation of the LiRA-style attack paradigm. This is why existing implementations \cite{carlini2022membership, ali2023membership, ye2022enhanced, zarifzadeh2023low} focus exclusively on TLC disclosures. Nonetheless, our theoretical results in Propositions \ref{Prop_1}--\ref{Prop_3} remain valid for arbitrary class sizes and continue to offer meaningful insights across all disclosure types.

To demonstrate the applicability of our framework in large-scale settings for current LiRA-style attacks \cite{carlini2022membership, ali2023membership, ye2022enhanced, zarifzadeh2023low}, we conduct experiments under TLC disclosure on CIFAR-100. We randomly select $N^{\mathrm{tr}}=4000$ images for data sets $\mathcal{D}^\mathrm{out}$ and $\mathcal{D}^{\mathrm{in}}$, and fine-tune a pre-trained ResNet-50 using SGD with a learning rate of $0.001$, momentum of $0.9$, and weight decay of $0.0005$ for $100$ epochs. Following the methodology in Sec. \ref{subsec_Dir}, we extract softmax outputs for class-$0$ samples and fit their distributions using the two-parameter Beta form in equation \eqref{Beta_in_out}, where $\gamma_0^z$ and $\overline{\gamma}_0^z$ represent the parameters for the true and aggregated non-true labels. As shown in Fig. \ref{fitting3}, the fitted distributions closely match the empirical curves, validating the flexibility of our modeling assumption for large $K$.

The results in Fig. \ref{fig_RCE_AU_EU_large_scale} further confirm that under large $K$, the trends under the TLC setting remain consistent with those observed in Fig. \ref{Fig_Adv_RCE}--Fig. \ref{Fig_Adv_EU} for TLC. This further supports the applicability of our theoretical framework to current LiRA-style attacks \cite{carlini2022membership, ali2023membership, ye2022enhanced, zarifzadeh2023low}, even in settings with a large number of classes.

\vspace{-5mm}
\bibliographystyle{IEEEtran}
\bibliography{MIA_analysis.bib}

\begin{thebibliography}{10}
\providecommand{\url}[1]{#1}
\csname url@samestyle\endcsname
\providecommand{\newblock}{\relax}
\providecommand{\bibinfo}[2]{#2}
\providecommand{\BIBentrySTDinterwordspacing}{\spaceskip=0pt\relax}
\providecommand{\BIBentryALTinterwordstretchfactor}{4}
\providecommand{\BIBentryALTinterwordspacing}{\spaceskip=\fontdimen2\font plus
\BIBentryALTinterwordstretchfactor\fontdimen3\font minus \fontdimen4\font\relax}
\providecommand{\BIBforeignlanguage}[2]{{%
\expandafter\ifx\csname l@#1\endcsname\relax
\typeout{** WARNING: IEEEtran.bst: No hyphenation pattern has been}%
\typeout{** loaded for the language `#1'. Using the pattern for}%
\typeout{** the default language instead.}%
\else
\language=\csname l@#1\endcsname
\fi
#2}}
\providecommand{\BIBdecl}{\relax}
\BIBdecl

\bibitem{shokri2017membership}
R.~Shokri, M.~Stronati, C.~Song, and V.~Shmatikov, ``Membership inference attacks against machine learning models,'' in \emph{Proc. IEEE Symp. Secur. Privacy}, May. 2017, pp. 3--18.

\bibitem{carlini2022membership}
N.~Carlini, S.~Chien, M.~Nasr, S.~Song, A.~Terzis, and F.~Tramer, ``Membership inference attacks from first principles,'' in \emph{Proc. IEEE Symp. Secur. Privacy}, May. 2022, pp. 1897--1914.

\bibitem{yeom2018privacy}
S.~Yeom, I.~Giacomelli, M.~Fredrikson, and S.~Jha, ``Privacy risk in machine learning: Analyzing the connection to overfitting,'' in \emph{Proc. IEEE Comput. Secur. Found. Symp.}, Jul. 2018, pp. 268--282.

\bibitem{jayaraman2019evaluating}
B.~Jayaraman and D.~Evans, ``Evaluating differentially private machine learning in practice,'' in \emph{Proc. 28th USENIX Security Symp.}, Aug. 2019, pp. 1895--1912.

\bibitem{chen2023overconfidence}
Z.~Chen and K.~Pattabiraman, ``Overconfidence is a dangerous thing: Mitigating membership inference attacks by enforcing less confident prediction,'' \emph{arXiv preprint arXiv:2307.01610}, Jul. 2023.

\bibitem{hullermeier2021aleatoric}
E.~H{\"u}llermeier and W.~Waegeman, ``Aleatoric and epistemic uncertainty in machine learning: An introduction to concepts and methods,'' \emph{Machine Learning}, vol. 110, no.~3, pp. 457--506, 2021.

\bibitem{guo2007sensitivity}
J.~Guo and X.~Du, ``Sensitivity analysis with mixture of epistemic and aleatory uncertainties,'' \emph{AIAA J.}, vol.~45, no.~9, pp. 2337--2349, Sep. 2007.

\bibitem{jia2019memguard}
J.~Jia, A.~Salem, M.~Backes, Y.~Zhang, and N.~Z. Gong, ``Memguard: Defending against black-box membership inference attacks via adversarial examples,'' in \emph{Proc. ACM SIGSAC Conf. Comput. Commun. Secur.}, Nov. 2019, pp. 259--274.

\bibitem{nasr2018machine}
M.~Nasr, R.~Shokri, and A.~Houmansadr, ``Machine learning with membership privacy using adversarial regularization,'' in \emph{Proc. ACM SIGSAC Conf. Comput. Commun. Secur.}, Oct. 2018, pp. 634--646.

\bibitem{Liu2025efficient}
Z.~Liu, W.~Jiang, F.~Zhou, and W.~Hongxin, ``Efficient membership inference attacks by bayesian neural network,'' \emph{arXiv preprint arXiv:2312.03262}, Mar. 2025.

\bibitem{salem2018ml}
A.~Salem, Y.~Zhang, M.~Humbert, P.~Berrang, M.~Fritz, and M.~Backes, ``{ML}-leaks: Model and data independent membership inference attacks and defenses on machine learning models,'' \emph{arXiv preprint arXiv:1806.01246}, Dec. 2018.

\bibitem{song2021systematic}
L.~Song and P.~Mittal, ``Systematic evaluation of privacy risks of machine learning models,'' in \emph{Proc. 30th USENIX Security Symp.}, Aug. 2021, pp. 2615--2632.

\bibitem{jayaraman2020revisiting}
B.~Jayaraman, L.~Wang, K.~Knipmeyer, Q.~Gu, and D.~Evans, ``Revisiting membership inference under realistic assumptions,'' \emph{arXiv preprint arXiv:2005.10881}, Jan. 2021.

\bibitem{liu2022membership}
Y.~Liu, Z.~Zhao, M.~Backes, and Y.~Zhang, ``Membership inference attacks by exploiting loss trajectory,'' in \emph{Proc. ACM SIGSAC Conf. Comput. Commun. Secur.}, Nov. 2022, pp. 2085--2098.

\bibitem{bertran2024scalable}
M.~Bertran, S.~Tang, A.~Roth, M.~Kearns, J.~H. Morgenstern, and S.~Z. Wu, ``Scalable membership inference attacks via quantile regression,'' \emph{Proc. Adv. Neural Inf. Process. Syst.}, vol.~36, 2024.

\bibitem{ali2023membership}
H.~Ali, A.~Qayyum, A.~Al-Fuqaha, and J.~Qadir, ``Membership inference attacks on {DNNs} using adversarial perturbations,'' \emph{arXiv preprint arXiv:2307.05193}, Jul. 2023.

\bibitem{ye2022enhanced}
J.~Ye, A.~Maddi, S.~K. Murakonda, V.~Bindschaedler, and R.~Shokri, ``Enhanced membership inference attacks against machine learning models,'' in \emph{Proc. ACM SIGSAC Conf. Comput. Commun. Secur.}, Nov. 2022, pp. 3093--3106.

\bibitem{zarifzadeh2023low}
S.~Zarifzadeh, P.~Liu, and R.~Shokri, ``Low-cost high-power membership inference attacks,'' \emph{arXiv preprint arXiv:2312.03262}, Jun. 2024.

\bibitem{li2021membership}
Z.~Li and Y.~Zhang, ``Membership leakage in label-only exposures,'' in \emph{Proc. ACM SIGSAC Conf. Comput. Commun. Secur.}, Nov. 2021, pp. 880--895.

\bibitem{choquette2021label}
C.~A. Choquette-Choo, F.~Tramer, N.~Carlini, and N.~Papernot, ``Label-only membership inference attacks,'' in \emph{Proc. Int. Conf. Mach. Learn.}, Jul. 2021, pp. 1964--1974.

\bibitem{vovk2005algorithmic}
V.~Vovk, A.~Gammerman, and G.~Shafer, \emph{Algorithmic learning in a random world}.\hskip 1em plus 0.5em minus 0.4em\relax Springer, 2005, vol.~29.

\bibitem{angelopoulos2021gentle}
A.~N. Angelopoulos and S.~Bates, ``A gentle introduction to conformal prediction and distribution-free uncertainty quantification,'' \emph{arXiv preprint arXiv:2107.07511}, Dec. 2022.

\bibitem{cohen2023calibrating}
K.~M. Cohen, S.~Park, O.~Simeone, and S.~Shamai, ``Calibrating {AI} models for wireless communications via conformal prediction,'' \emph{IEEE Trans. Cogn. Commun. Netw.}, vol.~1, pp. 296--312, Sep. 2023.

\bibitem{truex2019demystifying}
S.~Truex, L.~Liu, M.~E. Gursoy, L.~Yu, and W.~Wei, ``Demystifying membership inference attacks in machine learning as a service,'' \emph{IEEE Trans. Serv. Comput.}, vol.~14, no.~6, pp. 2073--2089, Feb. 2019.

\bibitem{tonni2020data}
S.~M. Tonni, D.~Vatsalan, F.~Farokhi, D.~Kaafar, Z.~Lu, and G.~Tangari, ``Data and model dependencies of membership inference attack,'' \emph{arXiv preprint arXiv:2002.06856}, Jul. 2020.

\bibitem{del2023bounding}
G.~Del~Grosso, G.~Pichler, C.~Palamidessi, and P.~Piantanida, ``Bounding information leakage in machine learning,'' \emph{Neurocomputing}, vol. 534, pp. 1--17, May. 2023.

\bibitem{aubinais2023fundamental}
E.~Aubinais, E.~Gassiat, and P.~Piantanida, ``Fundamental limits of membership inference attacks on machine learning models,'' \emph{arXiv preprint arXiv:2310.13786}, Jun. 2024.

\bibitem{kulynych2019disparate}
B.~Kulynych, M.~Yaghini, G.~Cherubin, M.~Veale, and C.~Troncoso, ``Disparate vulnerability to membership inference attacks,'' \emph{arXiv preprint arXiv:1906.00389}, Sep. 2021.

\bibitem{long2018understanding}
Y.~Long, V.~Bindschaedler, L.~Wang, D.~Bu, X.~Wang, H.~Tang, C.~A. Gunter, and K.~Chen, ``Understanding membership inferences on well-generalized learning models,'' \emph{arXiv preprint arXiv:1802.04889}, Feb. 2018.

\bibitem{dwork2014algorithmic}
C.~Dwork and A.~Roth, ``The algorithmic foundations of differential privacy,'' \emph{Found. Trends. Theor. Comput. Sci.}, vol.~9, no. 3--4, pp. 211--407, 2014.

\bibitem{wang2018stealing}
B.~Wang and N.~Z. Gong, ``Stealing hyperparameters in machine learning,'' in \emph{Proc. IEEE Symp. Secur. Privacy}, May. 2018, pp. 36--52.

\bibitem{murphy2012}
Y.~Polyanskiy and Y.~Wu, \emph{Information Theory From Coding to Learning}.\hskip 1em plus 0.5em minus 0.4em\relax Cambridge, U.K.: Cambridge Univ. Press, 2025.

\bibitem{jobic2023federated}
P.~Jobic, M.~Haddouche, and B.~Guedj, ``Federated learning with nonvacuous generalisation bounds,'' \emph{arXiv preprint arXiv:2310.11203}, Oct. 2023.

\bibitem{guo2017calibration}
C.~Guo, G.~Pleiss, Y.~Sun, and K.~Q. Weinberger, ``On calibration of modern neural networks,'' in \emph{Proc. Int. Conf. Mach. Learn.}, Aug. 2017, pp. 1321--1330.

\bibitem{kotz2004continuous}
S.~Kotz, N.~Balakrishnan, and N.~L. Johnson, \emph{Continuous Multivariate Distributions, Models and Applications}.\hskip 1em plus 0.5em minus 0.4em\relax Hoboken, NJ, USA: Wiley, 2004, vol.~1.

\bibitem{atchison1980logistic}
J.~Atchison and S.~M. Shen, ``Logistic-normal distributions: Some properties and uses,'' \emph{Biometrika}, vol.~67, no.~2, pp. 261--272, 1980.

\bibitem{bishop2006pattern}
C.~M. Bishop and N.~M. Nasrabadi, \emph{Pattern recognition and machine learning}.\hskip 1em plus 0.5em minus 0.4em\relax New York, NY, USA: Springer, 2006.

\bibitem{goodfellow2016deep}
I.~Goodfellow, Y.~Bengio, and A.~Courville, \emph{Deep learning}.\hskip 1em plus 0.5em minus 0.4em\relax Cambridge, MA, USA: MIT press, 2016.

\bibitem{valdenegro2022deeper}
M.~Valdenegro-Toro and D.~S. Mori, ``A deeper look into aleatoric and epistemic uncertainty disentanglement,'' in \emph{Proc. IEEE/CVF Conf. Comput. Vis. Pattern Recognit. Workshops}, Jun. 2022, pp. 1508--1516.

\bibitem{becker2022evaluating}
A.~Becker and T.~Liebig, ``Evaluating machine unlearning via epistemic uncertainty,'' \emph{arXiv preprint arXiv:2208.10836}, Sep. 2022.

\bibitem{youn2023randomized}
Y.~Youn, Z.~Hu, J.~Ziani, and J.~Abernethy, ``Randomized quantization is all you need for differential privacy in federated learning,'' \emph{arXiv preprint arXiv:2306.11913}, Jun. 2023.

\bibitem{liu2020privacy}
D.~Liu and O.~Simeone, ``Privacy for free: Wireless federated learning via uncoded transmission with adaptive power control,'' \emph{IEEE J. Sel. Areas Commun.}, vol.~39, no.~1, pp. 170--185, Nov. 2020.

\bibitem{papernot2021tempered}
N.~Papernot, A.~Thakurta, S.~Song, S.~Chien, and {\'U}.~Erlingsson, ``Tempered sigmoid activations for deep learning with differential privacy,'' in \emph{Proc. AAAI Conf. Artif. Intell.}, vol.~35, no.~10, Feb. 2021, pp. 9312--9321.

\bibitem{zecchin2024generalization}
M.~Zecchin, S.~Park, O.~Simeone, and F.~Hellstr{\"o}m, ``Generalization and informativeness of conformal prediction,'' \emph{arXiv preprint arXiv:2401.11810}, Jan. 2024.

\bibitem{byrd1995limited}
R.~H. Byrd, P.~Lu, J.~Nocedal, and C.~Zhu, ``A limited memory algorithm for bound constrained optimization,'' \emph{SIAM J. Sci. Comput.}, vol.~16, no.~5, pp. 1190--1208, 1995.

\bibitem{lin2016dirichlet}
J.~Lin, ``On the {Dirichlet} distribution,'' Master's thesis, Dept. Math. Statist., Queen's Univ., Kingston, ON, Canada, Sep. 2016.

\bibitem{cohen1998comparisons}
J.~Cohen, J.~H. Kempermann, and G.~Zbaganu, \emph{Comparisons of stochastic matrices with applications in information theory, statistics, economics and population}.\hskip 1em plus 0.5em minus 0.4em\relax Boston, MA, USA: Springer, 1998.

\bibitem{he2024information}
H.~He, C.~L. Yu, and Z.~Goldfeld, ``Information-theoretic generalization bounds for deep neural networks,'' \emph{arXiv preprint arXiv:2404.03176}, Apr. 2024.

\bibitem{kaya2020effectiveness}
Y.~Kaya, S.~Hong, and T.~Dumitras, ``On the effectiveness of regularization against membership inference attacks,'' \emph{arXiv preprint arXiv:2006.05336}, Jun. 2020.

\bibitem{wang2020against}
Y.~Wang, C.~Wang, Z.~Wang, S.~Zhou, H.~Liu, J.~Bi, C.~Ding, and S.~Rajasekaran, ``Against membership inference attack: Pruning is all you need,'' \emph{arXiv preprint arXiv:2008.13578}, Jul. 2021.

\bibitem{zade2025automatic}
S.~Z. Zade, Y.~Qiang, X.~Zhou, H.~Zhu, M.~A. Roshani, P.~Khanduri, and D.~Zhu, ``Automatic calibration for membership inference attack on large language models,'' \emph{arXiv preprint arXiv:2505.03392}, May. 2025.

\bibitem{wang2019miasec}
C.~Wang, G.~Liu, H.~Huang, W.~Feng, K.~Peng, and L.~Wang, ``{MIASec}: Enabling data indistinguishability against membership inference attacks in {MLaaS},'' \emph{IEEE Trans. Sustain. Comput.}, vol.~5, no.~3, pp. 365--376, Jul. 2019.

\bibitem{yang2023purifier}
Z.~Yang, L.~Wang, D.~Yang, J.~Wan, Z.~Zhao, E.-C. Chang, F.~Zhang, and K.~Ren, ``Purifier: Defending data inference attacks via transforming confidence scores,'' in \emph{Proc. AAAI Conf. Artif. Intell.}, vol.~37, no.~9, Feb. 2023, pp. 10\,871--10\,879.

\bibitem{hu2022defending}
L.~Hu, J.~Li, G.~Lin, S.~Peng, Z.~Zhang, Y.~Zhang, and C.~Dong, ``Defending against membership inference attacks with high utility by gan,'' \emph{IEEE Trans. Dependable Secure Comput.}, vol.~20, no.~3, pp. 2144--2157, May. 2022.

\bibitem{chzhen2021set}
E.~Chzhen, C.~Denis, M.~Hebiri, and T.~Lorieul, ``Set-valued classification--overview via a unified framework,'' \emph{arXiv preprint arXiv:2102.12318}, Feb. 2021.

\bibitem{devetyarov2010prediction}
D.~Devetyarov and I.~Nouretdinov, ``Prediction with confidence based on a random forest classifier,'' in \emph{Proc. IFIP Int. Conf. Artif. Intell. Appl. Innov.}\hskip 1em plus 0.5em minus 0.4em\relax Cham, Switzerland: Springer, Oct. 2010, pp. 37--44.

\end{thebibliography}
\end{document}